\documentclass[conference]{IEEEtran}
\IEEEoverridecommandlockouts
\usepackage{cite}
\usepackage{amsmath,amssymb,amsfonts}
\usepackage{algorithmic}
\usepackage{graphicx}
\usepackage{textcomp}
\usepackage{xcolor}
\def\BibTeX{{\rm B\kern-.05em{\sc i\kern-.025em b}\kern-.08em
    T\kern-.1667em\lower.7ex\hbox{E}\kern-.125emX}}
\usepackage{listings}
\usepackage{amsthm}

\usepackage{mathtools,array}
\usepackage{proof}
\usepackage{mathpartir}

\usepackage{changepage}
\usepackage{hyperref}

\usepackage{xcolor}
\usepackage{tabularray}
\usepackage{pdflscape}

\usepackage{adjustbox}

\usepackage{upquote}
\usepackage{etoolbox} 

\robustify{\texttt}
\let\originaltexttt\texttt

\begingroup
\catcode`'=\active
\catcode``=\active
\makeatletter
\renewrobustcmd{\texttt}[1]{%
   {%
   \everyeof{\noexpand}\endlinechar-1
   \expandafter\catcode\string``=\active
   \expandafter\catcode\string`'=\active
   \let'\textquotesingle
   \let`\textasciigrave
   \ifx\encodingdefault\upquote@OTone
    \ifx\ttdefault\upquote@cmtt
     \def'{\char13 }\def`{\char18 }%
    \fi
   \fi
   \scantokens{\originaltexttt{#1}}%
   }%
}%
\endgroup

\usepackage{listings}

\lstdefinestyle{listing0}{
  basicstyle=\ttfamily\footnotesize,
  moredelim  = [is][\color{blue}]{<@}{@>},
  moredelim  = [is][\bfseries]{/@}{@/},
  columns=fixed,
}

\theoremstyle{definition}
\newtheorem{definition}{Definition}
\newtheorem{remark}[definition]{Remark}
\newtheorem{theorem}[definition]{Theorem}
\newtheorem{lemma}[definition]{Lemma}
\newtheorem{example}[definition]{Example}

\newcommand{\inbrac}{\mathbin{{-}\hspace{-.22em}{[}}}
\newcommand{\outbrac}{\mathbin{{]}\hspace{-.21em}{\to}}}

\usepackage{supertabular}

\newenvironment{grammar}[2]
{\begin{supertabular}{@{\qquad}>{$}l<{$}@{\qquad}l@{}}
    \multicolumn{1}{@{}l@{}}{$#1$}&\multicolumn{1}{l@{}}{\hspace{-2em}#2}\\}
  {\end{supertabular}}

\usepackage{tikz}
\usetikzlibrary{arrows.meta}
\tikzset{>={latex}} 

\newcommand{\boxednode}[1]{{\footnotesize \tikz[baseline=(char.base)]{
		\node[shape=rectangle,
		semithick,
		draw,
		inner sep=3pt,
		rounded corners=4pt,
		minimum size=14pt] (char) {#1};}}}

\newcolumntype{L}{>{$}l<{$}} 

\begin{document}

\title{Tamgram: A Frontend for Large-scale Protocol Modeling in Tamarin
}

\author{Di Long Li\\
 \textit{The Australian National University}\\
 Canberra, Australia
 \and
Jim de Groot\\
 \textit{The Australian National University}\\
 Canberra, Australia
 \and
Alwen Tiu\\
 \textit{The Australian National University}\\
 Canberra, Australia}

\maketitle

\thispagestyle{plain}
\pagestyle{plain}

\begin{abstract}
Automated security protocol verifiers such as ProVerif and Tamarin have been increasingly applied to verify large scale complex real-world protocols. While their ability to automate difficult reasoning processes required to handle protocols at that scale is impressive, there remains a gap in the modeling languages used.
In particular, providing support for writing and maintaining large protocol specifications.
This work attempts to fill this gap by introducing a high-level protocol modeling language, called Tamgram,
with a formal semantics that can be translated to the
multiset rewriting semantics of Tamarin.
Tamgram supports writing native Tamarin code directly,
but also allows for easier structuring of large specifications through various high-level constructs,
in particular those needed to manipulate states in protocols.
We prove the soundness and the completeness of Tamgram with respect
to the trace semantics of Tamarin, discuss different translation strategies,
and identify an optimal strategy that yields performance comparable
to manually coded Tamarin specifications.
Finally we show the practicality of Tamgram with a set of small case studies
and one large scale case study.

\end{abstract}


\section{Introduction}

Fully automated and semi-automated protocol verifiers
have become a critical part of modern security
protocols analysis
as a comprehensive and fully manual analysis of their specifications become prohibitively expensive from the growing
set of requirements, constraints, and use cases. 

Tamarin~\cite{MeierSCB13} is a widely used tool for specifying and proving security properties of protocols. 
It is based on multiset rewriting rules, which allow for a precise encoding of stateful protocols.
This is of particular interest to us
as many widely used protocols, e.g.~WPA2~\cite{wpa2}, TLS~\cite{tls},
make use of states. 
Modeling these in other provers
such as ProVerif~\cite{proverif} may yield false attacks~\cite{sapic}.
Additionally, multiset rewriting rules
allow us to encode very precisely
interactions between concurrent processes,
such as resource
sharing and synchronisation,
which are difficult to capture as precisely
in other higher level languages based on process calculi,
such as variants of the applied pi-calculus~\cite{AbadiBF18}.

While Tamarin has been used
in various large scale case studies~\cite{BasinCDS22},
during our use of Tamarin for a case study
of a similar scale,
we encountered difficulty
in specifying a large
system in a modular manner.
We observed various workarounds
addressing similar difficulties
in existing case studies (see Section~\ref{sec:problems} for a discussion on these difficulties),
which made it more difficult
to understand the specifications.
These difficulties include
the manual handling of namespaces,
and the (often purely textual) macro expansion systems used,
which makes it hard to keep track of the scoping of names, as the macro systems
are not aware of the lexical scoping rules of Tamarin.

\subsection{Our contributions}

We present Tamgram,
a high-level frontend language to Tamarin which
addresses the usability issues by
introducing features such as
a Meta Language (ML) \cite{SML1997definition}-style module system,
user-defined predicate symbols,
additional let-binding sites,
an extensive support of hygienic macros
to maximize
code reuse as a first-class feature,
and
a syntax for processes which
allows for a specification
of complex process graphs
with an easy-to-use process-local
memory manipulation.
The translation procedure from Tamgram to Tamarin is in essence
a formal encoding of
the usual ad-hoc manual translation
of protocol control flows seen
in existing case studies.

The provision of 
programming constructs such as
(a restricted form of)
while loops and
if-then-else
branching
are important as they arise
naturally in specifications of
protocols, yet are highly error prone
to model manually in Tamarin rules
especially in presence
of sequential composition.

Tamgram distinguishes
itself from other Tamarin frontend languages
by using Tamarin rules as a core part
of the syntax,
allowing access to
almost all Tamarin features
(fine-tuning of parameters
and specification of heuristics
within same file).
During our experimentation,
we also identified optimization
techniques
that could benefit
similar work.

Finally, we formally show that
there is a close mapping
between a trace in Tamgram
and a trace in Tamarin of the translated system.
The formal result gives users assurance
that the intuitive mental model of
a Tamgram process is reliable,
and results proven by Tamarin
reflects accurately what
was specified in the Tamgram files.

\subsection{Related work}

The closest to our work is perhaps SAPIC (Stateful Applied PI-Calulus)\cite{sapic}, a frontend language to Tamarin that 
allows users to specify protocols in a dialect of
the applied pi-calculus with support for global state manipulation.
We revisit SAPIC in depth in Section~\ref{sec:compare-to-sapic}
after introduction of Tamgram
features.

SAPIC+\cite{sapic-plus} gives a unified interface language
targeting multiple solvers, namely
ProVerif, Tamarin and DeepSec~\cite{deepsec}.
The dialect of applied pi-calculus
used by SAPIC+ is similar
to SAPIC, but has several extensions,
such as failable let binding
with pattern matching.

A different direction in the design of a frontend language
for Tamarin is to support familiar informal notation often used in the security protocol literature,
the so-called ``Alice and Bob'' translation~\cite{BasinKRS15}.
This work, as far as we know, does not aim to support large scale protocol modeling. 

ProVerif \cite{proverif} is another widely used and
efficient protocol verifier. It uses the applied pi-calculus,
which excels in modeling
large scale and complex
protocols.
Its prover engine is based on Horn clauses, which makes encoding of states
not immediately obvious as its prover is essentially a classical first-order solver for which facts and clauses are persistent by default. 
However, there have been extensions to ProVerify that support encoding of states~\cite{ArapinisPRR14} and other features such as restrictions on states and the ability to use lemmas~\cite{proverif-overhaul-2022}.

\section{Background}

Since Tamgram heavily incorporates Tamarin syntax
and semantics, we briefly recall Tamarin
and some of its formal definitions.

\subsection{Overview of Tamarin}

Tamarin is based on multiset rewriting.
A \emph{multiset rewrite} (MSR) \emph{rule}
is a triple denoted by $l \inbrac a \outbrac r$,
where $l$, $a$, $r$ are each a sequence of multisets.
We name $l$ the premise, $a$ the action or label,
and $r$ the result or consequent.
A \emph{MSR system} is
a set of MSR rules.
The ``execution'' of a MSR system is then
defined as a labeled transition system,
where intuitively $l$ describes the accepted
state (state must contain all terms in $l$),
$a$ is the label,
and $r$ describes the modification.
(See Section~\ref{sec:semantics-tamarin}.)

In security protocol analysis, the properties
we want are often quantified over all traces
of execution,
e.g.~``for all traces, some secret key is not leaked.''
To prove such a property, Tamarin typically tries
to find a counterexample
by assuming the negation of the property (``key is leaked''),
then searching backwards.
If such a trace is possible then we have found an attack, and if not then we have a security guarantee.

The MSR system as used in Tamarin is ultimately a specification of a search problem 
so we need to be aware of the verification performance
when designing the translation procedure,
since unnecessary complexity of the translated
code or introduction of too many possible paths
can greatly increase verification time.

\subsection{Labeled operational semantics of Tamarin}\label{sec:semantics-tamarin}

  We sketch the labeled operational semantics of Tamarin here,
  using the symbols and operators
  from~\cite{tamarin-phd-thesis}.
  We assume that the reader is familiar with Tamarin syntax,
  so we shall not define what each construct of Tamarin means. 

\begin{definition} \
    \begin{itemize}
        \item Given a multiset $S$, $lfacts(S)$
        denotes the list of all
        linear facts in $S$,
        and $pfacts(S)$ denotes the list of persistent facts in $S$.
        \item Given a list $l$, we write $set(l)$ and $mset(l)$
        for the collection of elements of $l$
        as a set and multiset, respectively.
        \item Given multisets $A$ and $B$,
        $A \backslash^\sharp B$ denotes
        the multiset of all elements in $A$
        which are not in $B$.
        \item Given multisets $A$ and $B$,
        $A \cup^\sharp B$
        denotes the multiset in
        which the linear facts can be
        partitioned into $lfacts(A)$ and $lfacts(B)$,
        and all members of the persistent
        facts are present in $A$ or $B$
    \end{itemize}
\end{definition}

We introduce a shorthand that combines the consideration of persistent and linear facts:

\begin{definition}
  For multisets $S$ and $S'$ we write
	$
		S \subseteq^+ S'
	$
	if
	\begin{align*}
		lfacts(S) &\subseteq^\sharp lfacts(S') \\
            \text{and}\quad set(pfacts(S)) &\subseteq set(pfacts(S'))
	\end{align*}
\end{definition}

To encode a protocol in a MSR system, one defines a set of rules that capture, among others, the intruder's capability (following the Dolev-Yao intruder model~\cite{DolevY83}), and protocol steps. 
Additionally, there is a built-in rule that can be used to generate fresh names (for encoding nonces, keys, etc):
\begin{align*}
  FRESH = [] \inbrac \outbrac [ Fr(x:fr) ]
\end{align*}
Given a MSR system $R$, its transition semantics $steps(R)$ can be defined as
\begin{align*}
  \inferrule[]
  { l \inbrac a \outbrac r
    \in ginsts(R \cup \{ FRESH \})\\
    l\subseteq^+ S
  }
  { (S,
    l\inbrac a \outbrac r,
    S')
    \in steps(R)
  }
\end{align*}
Here we denote by $ginsts(R)$ the set of instances of the MSR system $R$
and $S' = (S\backslash^\sharp lfacts(l)) \cup^\sharp mset(r)$.

\begin{definition}
  Let $Sys$ be a set of rewrite rules modeling a protocol. 
  An \emph{execution trace} in Tamarin is a sequence alternating between
  multisets of facts 
  and ground instances of multiset rewriting rules:
  $S_0, l_0 \inbrac a_0 \outbrac r_0,
  S_1, \dots
  $,
  where
  $S_i, l_i \inbrac a_i \outbrac r_i,
  S_{i+1} \in steps(Sys)$.
\end{definition}

\section{Problems in combining MSR rules and the applied-pi calculus}

Since mature tooling for the applied-pi calculus already exists, and can be applied to large scale modelling of protocols, as a first step towards our language design, we explored the possibility of combining MSR rules with the applied-pi calculus syntax and semantics. However, in the end our conclusion was that a new syntax is desirable, as we explain next.  

\textit{\textbf{Loops}}:
Complex protocols often
have loops for retrying
a stage due to
timeouts, partial errors, etc.
These are straightforward
to express in an imperative
language as conditional loops
(using some hypothetical syntax):

\begin{lstlisting}[style=listing0]
process main =
  ...
  var stage1_succeeded = false;
  var counter = 0;
  while counter <= 5 {
    // stage1
    ...
    counter += 1;
  }
  if stage1_succeeded then {
    ...
  } else {
    ...
  }
\end{lstlisting}

In the applied-pi calculus, one would model this either using recursion or replication (i.e., the !-operator); but the former provides a better syntax for expressing scoping and process parameterisation. 
This requires breaking
the process into many subprocesses,
passing the states $v0, v1, \dots$ through
either global states or as
arguments to the process invocation:

\begin{lstlisting}[style=listing0]
// global state
cell stage1_succeeded = false;

let after_stage1(v0, v1, ...) =
  if stage1_succeeded then {
    ...
  } else {
    ...
  }

let stage1(v0, v1, ..., counter) =
  if counter <= 5 then {
    // stage1
    ...
    if successful then {
      stage1_succeeded := true;
    } else {
      stage1(v0, v1, ..., counter + 1)
    }
  }

let main =
    ...
    stage1(v0, v1, ..., 0);
    after_stage1(v0, v1, ...)
\end{lstlisting}

\textit{\textbf{Sequential composition and MSR rules:}} 
Another programming construct that is useful in programming and specification languages is the sequential composition. The applied-pi calculus, just as its predecessor the pi-calculus, is unusual in the sense that the calculus does not support a full sequential composition (between processes), and only allows a limited form of sequential behaviour expressed using action prefixes. 
Extending the (applied-)pi calculus with a proper sequential composition is known to be problematic (see, e.g., \cite{GehrkeR97}), essentially due to the interaction between name scoping (bound input names and scope extrusion) with sequential composition, that breaks the associativity of sequential composition. Adding MSR rules to the mix is likely going to increase the complexity of managing a sound semantics. As a side note, we noticed that even in the Tamarin manual~\cite{tamarin-manual},  users are advised against mixing MSR rules with the applied-pi-based syntax in SAPIC.

\textit{\textbf{Overlap in functionalities}}:
Some core applied-pi primitives, such
as $in$, $out$ and $event$
are easily expressible in MSR rules
with the corresponding special
predicates $In$, $Out$, and the action
field of a rule.
From a user's perspective,
the syntactical redundancy requires
making an unclear choice.
From a language design perspective,
the applied-pi syntax is too heavy
if we already have MSR rules
at our disposal.

Overall, this motivates
us to create an imperative style
language which does not
contain special syntax
that are easily
accomodated by MSR rules.

\section{Common modeling techniques and problems}
\label{sec:problems}
We first refer to the
EMVerify case study \cite{emverify} as it
demonstrates the core issues Tamgram
attempts to address, while not being
overly
complex compared to other case studies.
We then reference some more complex
techniques used in the WPA2 case study \cite{wpa2-tamarin}
to highlight some further insufficiency.
The code snippets quoted
are reformatted and truncated when
appropriate.

\subsection{EMVerify}
\label{sec:emverify}

Let us take the role of a
reviewer navigating through the Tamarin models.
We start with \texttt{Contactless.spthy}
as it is the template file which yields
all the concrete ``variants,''
where each variant
describes a particular setup.
Suppose we wish to examine how a terminal
is modeled. We would begin
with the following rule:
\begin{lstlisting}[style=listing0]
rule Terminal_Sends_GPO:
    let ... in
    [ Fr(~UN), !Value($amount, value) ]
  --[ OneTerminal(), Role($Terminal, 'Terminal') ]->
    [ Out(<'GET_PROCESSING_OPTIONS', PDOL>),
      <@Terminal_Sent_GPO@>($Terminal, PDOL) ]
\end{lstlisting}
We now analyse the rule:
\begin{itemize}
  \item
  The rule generates a
  fresh value \texttt{UN} and accesses
  a previously set value stored
  via the persistent fact \texttt{!Value(...)}
  
  \item
  The rule then labels this transition with \texttt{OneTerminal()}
  and \texttt{Role(...)}
  
  \item
  Finally, the rule outputs \texttt{PDOL} to attacker via
  the special fact \texttt{Out},
  and we are left with \texttt{Terminal\_Sent\_GPO}
\end{itemize}

Based on prior experience, we guess that \texttt{Terminal\_Sent\_GPO}
is a state fact that stores the current state/context
(or process memory so to speak).
To confirm this and see which are the intended
following rules, we search for rules which
consume this fact:
\begin{lstlisting}[style=listing0]
rule Terminal_Sends_ReadRecord:
    [ <@Terminal_Sent_GPO@>($Terminal, PDOL),
      In(<AIP, 'AFL'>) ]
-->[ Out(<'READ_RECORD', 'AFL'>),
      <@Terminal_Sent_ReadRecord@>($Terminal, PDOL, AIP) ]
\end{lstlisting}
We follow the control flow again and see
a branching in control flow:
\begin{lstlisting}[style=listing0]
rule Terminal_Receives_Records_SDA:
    let ... in
    [ <@Terminal_Sent_ReadRecord@>($Terminal, PDOL, AIP),      
      In(records), !IssuingCA($Bank, $CA),
      !CertCA($CA, <<...>, sign1>) ]
  --[ ... ]->
    [ <@Terminal_Ready_For_CVM@>($Terminal, ~PAN, $Bank,
    $CA, PDOL, AIP, pubkBank, 'Null', CVM) ]
        
rule Terminal_Receives_Records_CDA:
    let ... in
    [ <@Terminal_Sent_ReadRecord@>($Terminal, PDOL, AIP),      
      In(records), !IssuingCA($Bank, $CA),
      !CertCA($CA, <<...>, sign1>) ]
  --[ ... ]->
    [ <@Terminal_Ready_For_CVM@>($Terminal, ~PAN, $Bank,
    $CA, PDOL, AIP, pubkBank, pubkCard, CVM) ]
        
rule Terminal_Receives_Records_DDA:
    let ... in
    [ <@Terminal_Sent_ReadRecord@>($Terminal, PDOL, AIP),      
      !IssuingCA($Bank, $CA), In(records),
      !CertCA($CA, <<...>, sign1>) ]
  --[ ... ]->
    [ <@Terminal_Ready_For_DDA@>($Terminal, ~PAN, $Bank,
    $CA, PDOL, AIP, pubkBank, pubkCard, CVM) ]
\end{lstlisting}
The process of understanding the overall
behavior demands careful bookkeeping of
where names appear, and how they are (intended to be) used.
While it
seems manageable in small cases,
we have already had to manually examine many details
for just 5 out of 63 rules.

Switching sides,
it is then not difficult to see the
obstacles faced by Tamarin users:
for each state fact,
the name must be uniquely picked when appropriate,
and the terms passed to each state fact
must be carefully checked manually
to ensure correct ordering and naming,
as Tamarin does not provide static checking
in this regard.

For instance, if we changed the order of the arguments
to \texttt{Terminal\_Ready\_For\_CVM},
the subsequent part of the protocol may not execute
because the premises no longer match.
This may cause Tamarin to conclude that no attacks were
found
without warning even if there is one easily discoverable,
simply because part of the protocol is unreachable.

In practice, Tamarin users address these issues with
``sanity check'' lemmas (proof goals which
serve as unit tests).
These lemmas demand Tamarin to prove that
some paths are reachable, for instance:
\begin{lstlisting}[style=listing0]
lemma executable: exists-trace
  "Ex Bank PAN t #i #j #k #l.
    i < j & //Card-Terminal agreement
    Running(PAN, 'Terminal', <...>)@i &
    Commit('Terminal', PAN, <...>)@j & ..."
\end{lstlisting}
But similar to unit tests in a software engineering context,
it is difficult to ascertain manually
if we have covered all paths,
even in absence of loops.

\subsection{WPA2}
\label{sec:wpa2}

We observed further points of friction
when reviewing the more complex
WPA2 case study model \cite{wpa2-tamarin},
where the textual macro processor m4 \cite{m4}
is used heavily to emulate process macros.
We begin with one of the starting rules
in \texttt{wpa2\_four\_handshake.m4}:
\begin{lstlisting}[style=listing0]
rule Auth_Snd_M1 [...]: let ... in
    [ AuthState(~authThreadID, 'INIT_R1_SA', <...>)
    , Fr(~ANonce), Fr(~messageID) ]
    --[ ... ]->
    [ AuthState(~authThreadID, 'PTK_START', <...>),
    ..., <@OutEnc@>(m1, ~authThreadID, ~messageID, Auth_Snd_M1, Auth) ]

OutEncRuleDataFrame(Auth_Snd_M1, Auth)
\end{lstlisting}
All facts seem to be related to the
handling of states with the exception
of \texttt{OutEnc}.
The fact carries further peculiarity:
names \texttt{Auth\_Snd\_M1} and \texttt{Auth}
are not bound in the rule, and no other rules
consume \texttt{OutEnc} at first glance.

By examining another file \texttt{encryption\_layer.m4i},
we recognize that a unique output
fact \texttt{Out\_...} is instantiated for each rule
by expanding \texttt{OutEnc} (\texttt{\$1}
refers to the first argument, etc):
\begin{lstlisting}[style=listing0]
define(<@OutEnc@>, <!<@Out_$4@>($1, $2, $3)!>)
\end{lstlisting}
\texttt{OutEncRuleDataFrame}
expands into one or two rules
that can consume the output fact,
depending on whether encryption
is mandatory
(specified as an optional third argument \texttt{\$1}
to \texttt{OutEncRuleDataFrame}):
\begin{lstlisting}[style=listing0]
define(OutEncRuleDataFrame,<!dnl
EncryptionRule($1, $2, kDataFrame)
ifelse($3, only_encrypted, , PlainRule($1)) !>)
\end{lstlisting}
A unique output encryption
rule is instantiated for each rule
by expanding \texttt{EncryptionRule}:
\begin{lstlisting}[style=listing0]
dnl $1 = RuleName, $2 = Auth or Supp
define(EncryptionRule, <!dnl
rule OutRule_Enc_$1 [...]: let ... in
  [ <@Out_$1@>(message, ~senderThreadID, ~messageID)
  , $2SenderPTK(~ptkID, ~senderThreadID, ...) ]
  --[ ... ]->
  [ <@Out@>(snenc(message, PTK, newNonce))
  , $2SenderPTK(~ptkID, ~senderThreadID, ...) ] !>)
\end{lstlisting}
Similarly for \texttt{PlainRule}:
\begin{lstlisting}[style=listing0]
define(PlainRule, <!dnl
rule OutRule_$1 [color=ffffff]:
	[ <@Out_$1@>(message, ~senderThreadID, ~messageID) ]
	--[ SendMessage(~senderThreadID, ~messageID) ]->
	[ <@Out@>(message) ] !>)
\end{lstlisting}
Although the final expanded output is arguably
easier to review,
the careful
naming and bookkeeping required
to emulate lexically scoped
syntactic macros
can discourage reuse of the Tamarin code.

Overall, the experience is similar to
the handling of machine code: once written, it becomes
difficult to consistently update and
to review, as that
essentially demands reverse engineering.

\section{Basic components of a Tamgram process}

We explain Tamgram process in three steps: first the
Tamgram syntax itself,
then we briefly discuss the intermediate representation 
based on control flow
graphs (CFGs),
and finally the
semantics defined
in terms of the CFG.
We see the syntax in action
when revisiting the two case studies
in Section~\ref{sec:revisiting-case-studies}.

\subsection{Syntax}

The syntax of Tamgram takes inspiration from
applied pi-calculus (from ProVerif) and Tamarin,
and is defined as follows.

\phantom{text}

\begin{grammar}{path\Coloneqq}{}
  name & \\
  name.path & \\
\end{grammar}

\begin{grammar}{lvar\Coloneqq}{}
  x & bitstring variable \\
  \#x & timepoint variable \\
\end{grammar}

\begin{grammar}{cell\Coloneqq}{}
  'x & \\
\end{grammar}

\begin{grammar}{marg\_marker\Coloneqq}{}
    \text{named} & require explicit naming during application \\
\end{grammar}

\begin{grammar}{marg\_spec\Coloneqq}{macro argument specification}
    marg\_marker \dots name & \\
\end{grammar}

\begin{grammar}{marg\Coloneqq}{macro argument}
    term & \\
    name \texttt{ is } term & named argument \\
    name \texttt{ is } . & name pruning shorthands \\
    name \texttt{ is } '. & (. rewrites to $name$) \\
\end{grammar}

\begin{grammar}{term\Coloneqq}{}
  x & bitstring variable \\
  \$x & public symbol \\
  ``s'' & public string literal \\
  \sim x & fresh variable \\
  x \colon type & typed variable \\
  path & value \\
  cell & cell \\
  \text{let $x$ = $term$ in} & let binding \\
  \text{let $f(marg\_spec \dots)$ =} & bitstring macro \\
  \text{  $term$ in $term$} & \\
  \text{let $f(marg\_spec \dots) \colon type$ =} & term macro \\
  \text{  $term$ in $term$} & \\
  path(\dots) & application \\
  \langle term, \dots \rangle & tuple \\
  \text{$'name$ := $term$} & cell assignment \\
  \text{$'name$ := .} & name pruning \\
                         & (. rewrites to $name$) \\
  \text{undef($'name$)} & undefine cell \\
  \text{$term$ as $name$} & pattern matching with\\
  & naming \\
  \text{$'name$ cas $term$} & cell pattern matching \\
  \text{All $lvar$ $\dots$ \textbf{.} term} & universally quantified \\
  & formula \\
  \text{Ex $lvar$ $\dots$ \textbf{.} term} & existentially quantified \\
  & formula \\
\end{grammar}

\begin{grammar}{pmarg\_marker\Coloneqq}{}
    \text{named} & require explicit naming \\
                 & during application \\
    \text{rw} & allow read-write access \\
              & (only for cell arguments) \\
\end{grammar}

\begin{grammar}{pmarg\_spec\Coloneqq}{proc. macro arg. spec.}
    pmarg\_marker \dots name & \\
\end{grammar}

\begin{grammar}{pmarg\Coloneqq}{proc. macro arg.}
    marg & \\
    'name \texttt{ is } term & named cell argument \\
    'name \texttt{ is } . & name pruning \\
    'name \texttt{ is } '. & (. rewrites to $name$) \\
\end{grammar}

\begin{grammar}{decl\Coloneqq}{top level declaration}
  \text{let }name = term & let binding \\
  \text{fun }name / \mathbf{N} & uninterpreted \\
                               & function \\
  \text{pred }name / \mathbf{N} & uninterpreted \\
                               & predicate \\
  \text{apred }name / \mathbf{N} & uninterpreted action \\
  & predicate \\
  \text{let $x$ = $term$} & binding \\
  \text{fun $f(marg\_spec \dots)$} & bitstring macro \\
  \text{  $term$} & \\
  \text{pred $f(marg\_spec \dots)$} & predicate macro \\
  \text{  $term$} & \\
  \text{apred $f(marg\_spec \dots)$} & action predicate \\
  \text{  $term$} & macro \\
  \text{let $f(marg\_spec \dots) \colon type$} & term macro \\
  \text{  $term$} & \\
  \text{process $name$ = $P$} & process \\
  \text{process $name(pmarg\_spec \dots)$} & process macro \\
  \text{  $P$} & \\
  \text{module $name$ \{ $decl$ \dots \}} & submodule \\
  \text{import $name$} & import top-level \\
                       & module \\
  \text{open $name$ \{ $decl$ \dots \}} & unpack module\\
  &  $name$ into current\\
  & lexical scope \\
  \text{include $name$ \{ $decl$ \dots \}} & inherit bindings from \\
  & module $name$ \\
  \text{module $name$ = $path$} & module alias \\
\end{grammar}

\begin{grammar}{P\Coloneqq}{process}
  \mathbf{0}                        & null process \\
  \text{let $x$ = $term$ in $P$} & let binding \\
  \text{let $f(marg\_spec \dots)$ =} & bitstring macro \\
  \text{  $term$ in $P$ } & \\
  \text{let $f(marg\_spec \dots) \colon type$ =} & macro \\
  \text{  $term$ in $P$ } & \\
  rule; P         & rule \\
  ``anno": rule; P         & annotated rule \\
  & (text ``anno'' persists\\
  & after translation) \\
  choice \{ Pscoped, \dots \}; P & non-deterministic \\
  & choice \\
  Pscoped; P & scoped \\
  \text{while $cond$ } \{ P \}; P & \\
  \text{loop} \{ P \}; P & \\
  \text{if $cond$ then } \{ P \} \text{ else } \{ P \}; P & \\
\end{grammar}

\begin{grammar}{cond\Coloneqq}{}
  \text{$'name$ cas $term$} \phantom{01234567890123456789} & \\
  \text{($'name$ cas $term$)} \phantom{01234567890123456789} & \\
  \text{not ($'name$ cas $term$) } \phantom{01234567890123456789} & \\
\end{grammar}

\begin{grammar}{Pscoped\Coloneqq}{scoped process}
\{ P \} \phantom{01234567890123456789} & \\
\end{grammar}

\begin{grammar}{rule\Coloneqq}{rule}
  \text{$ruleL \to ruleR$} & unlabeled \\
  \text{$ruleL - ruleAR$} & labeled \\
\end{grammar}

\begin{grammar}{ruleL\Coloneqq}{left field}
  {[term,\dots]} & \\
\end{grammar}

\begin{grammar}{ruleAR\Coloneqq}{action and right field}
  {[term, \dots \outbrac} ruleR & \\
  \text{let $x$ = $term$ in $ruleAR$} & \\
  \text{let $f(\dots)$ = $term$ in $ruleAR$} & bitstring macro \\
  \text{let $f(\dots) \colon type$ = $term$ in} & macro \\
  \text{$ruleAR$} &  \\
\end{grammar}

\begin{grammar}{ruleR\Coloneqq}{right field}
{[term, \dots]} & \\
\text{let $x$ = $term$ in $ruleR$} & \\
\text{let $f(\dots)$ = $term$ in $ruleR$} & bitstring macro \\
\text{let $f(\dots) \colon type$ = $term$ in $ruleR$} & macro \\
\text{$ruleR$} & \\
\end{grammar}

Further syntactic restrictions:

\begin{itemize}
  \item
  $schevars(ruleL) \supseteq schevars(ruleAR)$ where $schevars(x)$ is the
  set of schematic variables
  referenced in $x$
  
  \item
  $Cell, St, StF, StB$ are not used anywhere (i.e., they are reserved symbols)
  
\end{itemize}

\subsection{Control flow graph representation}

The semantics of Tamgram is defined not directly on the process syntax presented earlier, but on an intermediate representation based on a control flow graph structure. This is to allow
flexibility of the high-level design,
and also to mirror Tamarin's semantics
more closely.

A Tamgram control flow graph (CFG) is a directed graph $\langle V, E \rangle$,
where $V$ is the set of vertices and $E$ the set of edges.
Each element of $V$ takes the form $(k,ru)$
where $k$ is a unique identifier (called its label) of the vertex,
and $ru$ is a MSR rule.
Since each vertex is identified by its label,
we can define $E$ in terms of the ids,
i.e., $E \subseteq \{(k,k') \mid (k,ru), (k,ru') \in V\}.$ 

We give an example of construction here:
\begin{example}
\begin{lstlisting}[style=listing0]
process A =
  [ In(x) ]->[ 'a := x ];
  choice {
    { [ 'a cas "1" ]->[ Out("2") ] };
    { [ 'a cas "A" ]->[ Out("B") ] };
  };
  []->[ Out("End") ]
\end{lstlisting}

The set $V$ then consists of:
\begin{center}
\begin{tabular}{LL}
	(0, & [ Fr(~pid) ]\rightarrow['pid := ~pid])\\
	(1, & [ In(x) ]\rightarrow[ 'a := x ])\\
	(2, & [ 'a \text{ cas } ``1'' ]\rightarrow[ Out(``2'') ])\\
	(3, & [ 'a \text{ cas } ``A'' ]\rightarrow[ Out(``B'') ])\\
	(4, & []\rightarrow[ Out(``End'') ])
\end{tabular}
\end{center}
where rule $0$ is automatically added
by Tamgram to model process initialization.
$E$ could then be visually represented as:

\begin{figure}[h]
	\centering
	\begin{tikzpicture}[scale=0.75]
		\node (Box0) at (0,2) {\boxednode{0}};
		\node (Box1) at (2,2) {\boxednode{1}};
		\node (Box2) at (4,2.5) {\boxednode{2}};
		\node (Box3) at (4,1.5) {\boxednode{3}};
		\node (Box4) at (6,2) {\boxednode{4}};
		
		\draw[semithick, ->] (Box0) to (Box1);
		\draw[semithick, ->] (Box1) to (Box2);
		\draw[semithick, ->] (Box1) to (Box3);
		\draw[semithick, ->] (Box2) to (Box4);
		\draw[semithick, ->] (Box3) to (Box4);
	\end{tikzpicture}
	\caption{Example toy CFG.}
	\label{fig:toy_cfg}
\end{figure}
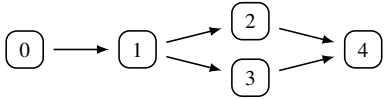
\end{example}

Overall, the CFG construction
matches the usual intuition of CFG.
We formally define
the CFG procedure in the technical report
corresponding to the paper.

\subsection{Labeled operational semantics of Tamgram}

Before we define the labelled transition semantics for Tamgram, we introduce 
some terminology and symbols used in the semantics. 
\begin{itemize}
    \item A {\em process memory} is a mapping from cells to cell-free terms. 
    \item Given a process memory $m$, 
    $\mathbb{D}(m)$ denotes its domain. 
    \item  Given a process memory $m$, we define a dereference function $deref(m, c) = t$ if $(c,t) \in m.$  This is extended homomorphically to a mapping from terms to cell-free terms. 
\item
Given a rule $r$, $undefs(r)$ denotes the set of facts of the form undef($'c$) in the right hand side of $r$.

\item Given a rule $r$, 
$defs(r) = \{ 'c \mapsto t \mid \text{ $'c := t$ is a fact in $r$}\}. $

\item
$ginsts(e)$ denotes set of
all ground instances of term $e$.
A ground instance of a term
only contains constants or cells as atoms.

\item Given a list of terms $r_0, \dots$,
$nostmt(r_0, \dots)$ (no statement)
yields $r'_0, \dots$
which is same as $r_0, \dots$
but without any
assignments (e.g.~$c := x$)
or undef,
and when provided with rule $l \inbrac a \outbrac r$,
yields
$l\inbrac a \outbrac nostmt(r)$
\end{itemize}
Additionally, we use the following sets:
\begin{itemize}
  \item $\mathcal{S}$ refers to the multiset of facts.
  
  \item $\mathcal{K}$ is a mapping from process ID to a label in graph
  representation of the process, serving as ``process counter''.
  
  \item $\mathcal{M}$ is a mapping from process ID to ``memory store,''
  each memory store is a mapping from cells to cell-free terms, serving as ``process memory.''
  We use this two layered mapping instead of single layer for easier specification
  of semantics.
\end{itemize}

We are now ready to define the labelled
operational semantics of Tamgram.

\begin{definition}
  The state-space of the labelled transition relation is
  the product 
  $(\mathcal{S} \times \mathcal{K} \times \mathcal{M})$.
  Transitions are labelled by ground instances of MSRs, i.e.~by
  elements of $ginsts(R\cup \{FRESH\})$.
  The transition relation
  $steps(Sys)
    \subseteq (\mathcal{S} \times \mathcal{K} \times \mathcal{M})
\times
ginsts(R\cup \{FRESH\})
\times
(\mathcal{S} \times \mathcal{K} \times \mathcal{M})$  
  is defined by the inference rules given in Figure~\ref{fig:inf-rules}.
  We detail the more complex definitions
  used by (\textsc{Rule}) below.
  
  Given a multiset $S$, process memory $m$,
  and a ground instance $l' \inbrac a' \outbrac r'$ of the MSR rule:
  \begin{itemize}
      \item
      We first remove dereferenced linear facts from $S$:\\
      $S_0 = S\backslash^\sharp lfacts(deref(m, l'))$
      \item
      then add the consequent facts as defined
      by $r'$ for the final multiset of facts:\\
      $S' = S_0 \cup^\sharp mset(deref(m, nostmt(r')))$
      \item
      Set of cells $cells_{old}$
      from $m$ which remain well defined
      after execution of $l' \inbrac a' \outbrac r'$
      but not overwritten
      is equal to
      $(\mathbb{D}(m)\backslash undefs(r'))\backslash \mathbb{D}(defs(r'))$
      \item
      Part of the memory $m_{old}$ to carry over
      from $m$ is then equal to
      $\{ (c, x) | c \in cells_{old} \land (c, x) \in m \}$
  \end{itemize}

\begin{figure*}
\centering
  $$
    (\textsc{Fresh})
    \inferrule[]
      { }
      {\langle (S, K, M), []\inbrac \outbrac [Fr(x)], (S, K, M) \rangle \\\\
        \text{where $x$ is a fresh name}}
  \qquad
    (\textsc{Start})
    \inferrule[]
      { (k, [Fr(id) ] \inbrac \outbrac ['pid := id ]) \in V}
      {\langle (S, K, M), [Fr(id)]\inbrac \outbrac [ 'pid := id ],\\ (S, K', M') \rangle \\\\
      \text{where $k \in roots(E)$} \\\\
      \text{$k' \in succ(E, k) $} \\\\
      \text{$K' = K \cup \{ id' \mapsto k' \}$,} \\\\
      \text{$M' = M \cup \{ id' \mapsto \{ 'pid \mapsto id' \}\}$}
      }
  $$
  $$
    (\textsc{Rule})
      \inferrule[]
  {(k, l \inbrac a \outbrac r) \in V \\
   l' \inbrac a' \outbrac r'
   \in ginsts(l \inbrac a \outbrac r)\\
   deref(m, l') \subseteq^+ S\\
   K = K' \cup \{ id \mapsto k \}\\
   M = M' \cup \{ id \mapsto m \}\\
   cu_r(l \inbrac a \outbrac r)\subseteq \mathbb{D}(m)
  }
  {\langle
    (S, K, M),
    l' \inbrac a' \outbrac r',\\
    (S',
    K' \cup \{ id \mapsto k' \},
    M' \cup \{ id \mapsto m_{old} \cup m_{new} \}
    )
  \rangle \\\\
  \text{where $k' \in succ(E, k)$ is picked non-deterministically}\\
  m_{new} = defs(r')
  }
  $$
  \caption{Inference rules for the labeled transition relation of a
           Tamgram system.}
  \label{fig:inf-rules}
\end{figure*}

\end{definition}

  Finally, we define the notion of a well-formed execution trace.

\begin{definition}
  An \emph{execution trace} in Tamgram is a sequence alternating between
  $\mathcal{S} \times \mathcal{K} \times \mathcal{C}$
  and ground instances of multiset rewriting rules:
  $(S_0, K_0, M_0), l_0 \inbrac a_0 \outbrac r_0,
  (S_1, K_1, M_1), \dots
  $,
  such that
  $(S_i, K_i, M_i), l_i \inbrac a_i \outbrac r_i,
  (S_{i+1}, K_{i+1}, M_{i+1})$ is in $steps(Sys)$
  for all $i$.
\end{definition}

\begin{definition} \label{def:well-formed-execution-trace}
  An execution trace is well-formed iff
  there is no invalid cell access
  e.g. reading cell $'c$ before it is defined.
\end{definition}

\begin{example}
Suppose a process
contains steps $[In(x)]\inbrac~\outbrac ['a := x, F(x)]$
and $[In(y)]\inbrac~\outbrac ['b := y, undef('a)]$,
then an example of a well-formed execution trace would be as follows
(we fulfill $In$ facts implicitly):
\begin{align*}
  (\emptyset, \emptyset, \emptyset),
  [Fr(\sim pid) ]\inbrac~\outbrac[ 'pid := pid],\\
  (\emptyset,
  \{ pid \mapsto 1 \},
  \{ pid \mapsto \{ 'pid \mapsto pid \} \}),\\
  [In(x)]\inbrac~\outbrac ['a := x, F(x) ],\\
  (\{ F(x) \},
  \{ pid \mapsto 2 \},
  \{ pid \mapsto \{ 'pid \mapsto pid, 'a \mapsto x \} \}),\\
  [In(y)]\inbrac~\outbrac ['b := y, undef('a)].\\
  (\{ F(x) \},
  \{ pid \mapsto 2 \},
  \{ pid \mapsto \{ 'pid \mapsto pid, 'a \mapsto x \} \})
\end{align*}
\end{example}

\section{Revisiting EMVerify and WPA2 in Tamgram}
\label{sec:revisiting-case-studies}

We note that while we rewrote
EMVerify fully in Tamgram,
as well as a
substantial part of WPA2 case study,
we only have comparable benchmark
results for EMVerify contactless
variants,
and discussion of WPA2
remains syntactical.
The reason is twofold.
Firstly,
both the EMVerify contact
variants and the WPA2 cases
require use of custom Tamarin oracles.
WPA2 in particular is also
heavily tuned via
fact annotations.
But it is unclear how these
should carry over to a Tamgram
model.
Secondly, Tamarin is transitioning
to a new tactic language for
ranking facts \cite{tamarin-tactics-lang},
and along with our plan
of extending Tamgram
to automate fact ranking,
translating existing oracles
becomes needlessly time consuming.

\textit{\textbf{State handling and composable process structure}}:
We begin by modeling the same terminal
from Section~\ref{sec:emverify}
in Tamgram.
We make use of Tamgram's main
state handling primitive:
process-local memory cells.
Cell names are prefixed by \text{\textquotesingle }
in Tamgram code,
and are dynamically scoped rather than lexically scoped.
A cell $c$ is assigned
a value $v$ via
the assignment
\texttt{'c := v} syntax,
or removed from process
memory via \texttt{undef(c)}.

We also make
use of the annotation syntax
$``anno": rule$ to help us refer
back to the original Tamarin model
by annotating the relevant steps
with the original rule names.
These annotations are preserved
after translation.

\begin{lstlisting}[style=listing0]
process Terminal =
  "<@Terminal_Sends_GPO@>":
    [ Fr(~UN),
      !Value($amount, value) ]
  --let ... in ... [ .... ]->
    [ Out(<"GET_PROCESSING_OPTIONS", PDOL>),
      'PDOL /@:=@/ PDOL, 'Role0 /@:=@/ $Terminal ];
  "<@Terminal_Sends_ReadRecord@>":
    [ 'Role0 cas $Terminal,
      In(<AIP, "AFL">) ]-->
    [ Out(<"READ_RECORD", "AFL">), 'AIP /@:=@/ AIP ];
  /@choice@/ {
    { "<@Terminal_Receives_Records_SDA@>":
        [ 'Role0 cas $Terminal,
          'AIP cas <"SDA", furtherData>,
          In(<...> as records), ... ]
      --[ ... ]->
        [ 'Role0 /@:=@/ $Terminal, 'PAN /@:=@/ PAN
        , ... ] };
    { "<@Terminal_Receives_Records_CDA@>":
        [ 'Role0 cas $Terminal,
          'AIP cas <"CDA", furtherData>,
          In(<...> as records), ... ]
      --[ ... ]->
        [ 'Role0 /@:=@/ $Terminal, ... ] };
    { "<@Terminal_Receives_Records_DDA@>":
        [ 'Role0 cas $Terminal,
          'AIP cas <"DDA", furtherData>,
          In(<...> as records), ... ]
      --[ ... ]->
        [ 'Role0 /@:=@/ $Terminal, ... ];
      "<@Terminal_Sends_InternalAuthenticate@>": ...;
      "<@Terminal_Receives_InternalAuthentic...@>":
        ... }; };
  ...
\end{lstlisting}

The introduction of cells comes from the observation
that state facts
are meaningfully packaging of
names which we wish to persist
past the lexical scope of a MSR rule.
And a process local mutable
memory cell satisfies this
criterion in an intuitive way.

Now a reviewer
no longer needs to hunt down
usage of facts laboriously,
as the relationships between
between rules are immediate via
sequential composition of steps,
and with \texttt{choice}
or other similar primitives
to denote divergence
of control flow.

As a user, since
we no longer need to derive our
own state predicates,
we can modify process structure
freely,
while enjoying the full expressive power
of MSR rules.

\textit{\textbf{Named arguments}}:
Although process-local cells
free us from the need to fill in
a state fact with a large
number of of arguments manually,
we inevitably need to
fill in a similarly complex fact
that cannot be simplified by using
cells. These might be action facts,
or shared
state between
multiple processes to model
concurrency.

As the WPA2 case study as an example,
we focus on the predicate
\texttt{AuthGTKState}, which constructs
a shared state fact between
multiple threads/processes,
and the action predicate
\texttt{AuthenticatorSendsInitialM3}.

\begin{lstlisting}[style=listing0]
rule Auth_Check_MIC_M2_Snd_M3:
  let ... in
  [ AuthState(...), ...
  , <@!AuthGTKState@>(~authID, stateIdentifier,
                  installedGTKData, shareGTKData,
                  ~pointerGTKState) ]
  --[ Eq(mic_m2, MIC(newPTK, m2))
    , <@AuthenticatorSendsInitialM3@>(~authThreadID,
                                  ~authID, ~PMK,
                                  ~suppThreadID,
                                  oldPTK, 
                                  ~ANonce, SNonce,
                                  ctr_plus_1)
    , ... ]->
  [ AuthState(...), ... ]
\end{lstlisting}

Both carry a considerable number of arguments
and usability suffers in the same way
to filling in state facts manually:
we may easily misplace arguments without
warning, yielding a different protocol model.

Tamgram provides guardrails 
for these cases as well by
introducing named (or labelled) arguments
seen commonly in a wide variety of programming
languages.
To utilize help from Tamgram,
we simply declare the predicates with named arguments
instead of just specifying the arity:
\begin{lstlisting}[style=listing0]
pred <@!AuthGTKState@> (named authID,
                    named state,
                    named installedGTKData,
                    named shareGTKData,
                    named pointerGTKState)

apred <@AuthenticatorSendsInitialM3@> (
  named authThreadID, named authID,
  named PMK,          named suppThreadID,
  named PTK,          named ANonce,
  named SNonce,       named ctr)
\end{lstlisting}

Usage of these predicates then requires
explicit unordered pairing of terms to
the argument label via the \texttt{is} keyword.
Since we often pair
a label with a variable or cell of the same name,
Tamgram provides name pruning syntax
\texttt{.} and \texttt{'.}
to denote such cases,
e.g. \texttt{a is '.} rewrites
to \texttt{a is 'a}.

\begin{lstlisting}[style=listing0]
"Auth_Check_MIC_M2_Snd_M3":
  [ ...
  , <@!AuthGTKState@>(authID /@is@/ ~authID,
                  state /@is@/ stateIdentifier,
                  installedGTKData /@is@/ .,
                  shareGTKData /@is@/ ...,
                  pointerGTKState /@is@/ .) ]
--let ... in ...
  [ Eq('mic_m2, MIC(newPTK, 'm2))
  , <@AuthenticatorSendsInitialM3@>(
      authThreadID /@is@/ '.,
      authID /@is@/ '.,
      PMK /@is@/ '.,
      suppThreadID /@is@/ '.,
      ...)
  , ... ]->[ ... ];
...;
\end{lstlisting}

This removes the need to scrutinize
over ordering of arguments entirely,
and also makes adding arguments to existing
predicates significantly easier.

\textit{\textbf{Module system}}:
In the WPA2 model,
all participants utilize
the same set
of encryption primitives with
same style of bookkeeping,
sharing the same ``encryption layer'' code so to speak.
To maximize code reuse,
the original Tamarin model
separates this common layer
out into \texttt{encryption\_layer.m4i}.
The code is then instantiated when appropriate
as seen in the snippets from Section~\ref{sec:wpa2}.

We noted that in order to
mimic a module namespace and to ensure
that the final expanded form does not
interfere with existing names,
a fair number of non-obvious mechanisms
were used,
such as magic string prefixes
and myriad internal macros.
This style requires careful consideration
of the main model,
and even more so should we wish to introduce
an additional module.

Tamgram addresses the difficulty
in packaging code by introducing
a simple ML-style module system.
As we are following
the original model,
we want to separate
the encryption layer from
the main model, but there
are also function symbols
that we wish to use across both modules.
We group these symbols in file
\texttt{fun\_symbols.tg}:

\begin{lstlisting}[style=listing0]
fun NULL() = "NULL"

// S denotes the successor function, GTK indicates that we are
// dealing with a group key
fun KDF/1
fun GTK/1
...
apred Neq/2
apred Eq/2

fun <@kNullPTK@>() = KDF(<NULL(), NULL(), NULL()>)
fun kNullGTK() = GTK(NULL())
...
\end{lstlisting}

Some of the names are term macros, e.g. \texttt{kNullPTK},
which help further tackle
the complexity of the model.

We then import the above
file (exposed as \texttt{Fun\_symbols}
in the module system)
in \texttt{encryption\_layer.tg}:

\begin{lstlisting}[style=listing0]
/@import@/ <@Fun_symbols@>
/@open@/ <@Fun_symbols@>

apred EnqueueMessage/2
apred SendMessage/2
...
\end{lstlisting}

In \texttt{wpa2\_four\_way\_handshake.tg},
we import both \texttt{Fun\_symbols}
and \texttt{Encryption\_layer},
and also define additional submodules
\texttt{Restrictions} and \texttt{Enc\_restrictions}.
We additionally introduce an alias \texttt{Enc} to
the encryption
layer module.

\begin{lstlisting}[style=listing0]
/@import@/ <@Fun_symbols@>
/@open@/ <@Fun_symbols@>
...
/@module@/ Restrictions = {
  restriction Neq =
      All x y #i. Neq(x, y) @ i ==> not(x = y)
  ...
}

/@import@/ <@Encryption_layer@>
/@module@/ <@Enc@> /@=@/ <@Encryption_layer@>

/@module@/ Enc_restrictions = {
  restriction MessagesAreSentInEnqueueOrder =
       All ... #i1 #j1 #i2 #j2.
       <@Enc@>.EnqueueMessage(...) @ i1 & ...
       <@Enc@>.SendMessage(...) @ i2 & ...
}
...
\end{lstlisting}

\textit{\textbf{Process macros}}:
Finally we use process
macros to provide a consistent
interface for participants
to manage the encryption layer.

Process macros differ from
term macros such as \texttt{kNullPTK}
in two main ways:
process macros always
operate in process syntax,
and process macros can treat
cell arguments differently from
term arguments. For instance,
term macros cannot make use of
the cell assignment (\texttt{:=}) syntax,
and as such any cell passed
to a term macro is always
semantically equivalent to
the underlying value.

We define the interfaces
\texttt{InEnc} and \texttt{OutEnc}
in \texttt{encryption\_layer.tg},
where we mark the cell \texttt{'message}
as \texttt{rw} to denote
we allow process macro
\texttt{InEnc} to write
back to the original process
through this cell.
Tamgram
also provides name pruning
syntax for
cell assignments similar
to the ones for named arguments:

\begin{lstlisting}[style=listing0]
process InEnc(named role,
              named /@rw@/ <@'message@>,
              named receiverThreadID,
              named receiverID) =
  choice {
    { "Receive_plaintext":
        [ In(<@message@>)
        , !ReceiverPTK(...) ]
      --[ ... ]->
        [ <@'message@> := /@.@/ ] };
    { "Receive_ciphertext":
        [ In(snenc(<@message@>, PTK, nonce))
        , !ReceiverPTK(...) ]
      --[ Neq(PTK, kNullPTK()), ... ]->
        [ <@'message@> := /@.@/ ] }; }
process OutEnc(named senderThreadID,
               named <@message@>,
               named frameTag) =
    [ Fr(~messageID) ]
  --[ EnqueueMessage(senderThreadID, messageID) ]->
    [ Queued(senderThreadID is .,
             messageID is .,
             <@message@> is .,
             frameTag is .) ]
\end{lstlisting}

Making use of the interfaces
in \texttt{wpa2\_four\_way\_handshake.tg}
is then as straightforward as using
other function symbols:

\begin{lstlisting}[style=listing0]
"Auth_Rcv_M2":
    [ 'state cas "PTK_START" ]->[];
  <@Enc.InEnc@>(
    role is Auth(),
    <@message@> is '.,
    receiverThreadID is 'authThreadID,
    receiverID is 'authID);
    [ <@'message@> cas <<'ctr, SNonce> as m2, mic_m2> ]
  --[ ... ]->[ ... ]
...
"Auth_Check_MIC_M2_Snd_M3":
    [ 'state cas "PTK_CALC_NEGOTIATING"
    , ... ]
  --let ... in ...
    [ ... ]->[ ... ];
  <@Enc.OutEnc@>(
    <@message@> is <'m3, MIC('newPTK, 'm3)>,
    senderThreadID is 'authThreadID,
    frameTag is kDataFrame())
\end{lstlisting}

Overall, process macros in combination
with the module system
provide a straightforward
way to
package away details
not
immediately relevant to the main model.

\section{Comparison to SAPIC} \label{sec:compare-to-sapic}

The most prominent difference
between Tamgram and SAPIC
is that SAPIC is integrated into
Tamarin, while Tamgram is
a separate compiler.
We elaborate upon other major differences
below. We assume
reader is familiar with the
usual syntax of a process in pi-calculus.

\textit{\textbf{Unit of steps}}:
Tamgram uses MSR rules as
the smallest unit of steps,
while the core langauge of SAPIC models
steps in its variant of applied pi-calculus.

\textit{\textbf{State handling}}:
In SAPIC, in the core langauge
there is only notion of global states:

\begin{grammar}{P, Q\Coloneqq}{stateful process}
  \text{insert $t1$, $t2$; $P$} & set state $t1$ to $t2$ \\
  \text{delete $t$; $P$} & delete state $t$ \\
  \text{lookup $t$ as $x$ in $P$ else $Q$; $P$} & read state $t$ \\
        & into variable $x$ \\
\end{grammar}

To store a local state $v$, we pick
a term $k$ that is unique to the process (usually
a process ID of sorts),
and then use the \texttt{insert k, v; P} syntax.
This lack of separation
makes error checking
concerning process local states
difficult to perform, e.g.
use of uninitialized states,
use of deleted states.
Some translation optimizations
discussed in Section~\ref{sec:translation-styles}
are also made complicated in
the general case.

\textit{\textbf{Modeling concurrency}}:
SAPIC provides syntax for modeling
a typical mutex:

\begin{grammar}{P, Q\Coloneqq}{stateful process}
  \text{lock $t$; $P$} & set lock on $t$ \\
  \text{unlock $t$; $P$} & remove lock on $t$ \\
\end{grammar}

In Tamgram we need to
specify the relevate predicates
and restrictions ourselves for lock.

Though as a trade-off,
in Tamgram we can also specify
concurrency primitives or data structures
not related to locks or lookup table entries
in MSR rules directly,
e.g. semaphore, queue, condition variables.

\textit{\textbf{Further abstractions}}:
SAPIC provides further features out of the box.
For instance,
translating using a ``enforcing local progress''
semantics,
where the set of traces considered
consist of only traces of processes
which have been reduced as far as possible,
and ``Isolated Execution Environments'' (IEEs)
or enclaves.

\textit{\textbf{Usability improvements}}:
Tamgram provides a module system
and named arguments
which make
models easier to package.
But these syntactic extensions
are easy to incorporate
into SAPIC,
and do not constitute fundamental differences.

Overall, for protocols which do not
make use of many local states,
and do not need the expressive
power of MSR rules,
SAPIC is likely more
straightforward and succinct.
On the other hand, Tamgram is designed
to work well with MSR rules,
and is designed
for processes with many local states
and control flows not easily expressible
in the pi process syntax.

\section{Translation into Tamarin}

The way we translate Tamgram into Tamarin
has a significant impact
on the verification speed.
Models handwritten
and fine-tuned by experts
are usually the fastest compared to
translated models~\cite{sapic-plus}.
But it is difficult to handcraft
a model for a complex protocol,
and hence we want Tamgram
to capture the ``expert'' style.

We start by exploring three naive ways of
translation, and then combine these into a ``hybrid'' translation which is closest to the ``expert'' style.

We only consider
CFGs that always yield
well-formed execution traces
(Definition~\ref{def:well-formed-execution-trace}).
The procedure which determines
if a given CFG has this property
is in essence
a different formulation of
live-variable analysis
and static-single assignment
\cite{appel1998modern}.
We relegate the details
and correctness proof
of this procedure
and the correctness proof of the overall
translation procedure
to the technical report.\footnote{Available in the supplementary material included in this submission.}

\subsection{Translation styles and considerations} \label{sec:translation-styles}

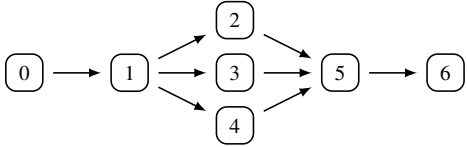
\begin{figure}[h]
	\centering
	\begin{tikzpicture}[scale=0.70]
		\node (Box0) at (0,2) {\boxednode{0}};
		\node (Box1) at (2,2) {\boxednode{1}};
		\node (Box2) at (4,3) {\boxednode{2}};
		\node (Box3) at (4,2) {\boxednode{3}};
		\node (Box4) at (4,1) {\boxednode{4}};
		\node (Box5) at (6,2) {\boxednode{5}};
		\node (Box6) at (8,2) {\boxednode{6}};
		
		\draw[semithick, ->] (Box0) to (Box1);
		\draw[semithick, ->] (Box1) to (Box2);
		\draw[semithick, ->] (Box1) to (Box3);
		\draw[semithick, ->] (Box1) to (Box4);
		\draw[semithick, ->] (Box2) to (Box5);
		\draw[semithick, ->] (Box3) to (Box5);
		\draw[semithick, ->] (Box4) to (Box5);
		\draw[semithick, ->] (Box5) to (Box6);
	\end{tikzpicture}
	\caption{Basic CFG.}
	\label{fig:basic_cfg}
\end{figure}

We demonstrate the naive translations
using CFG as shown in Figure~\ref{fig:basic_cfg}.
Throughout the subsections,
we use $St(k, \dots)$ to denote our state facts,
where
$k$ denotes the rule number, i.e.~$k \in \mathbb{D}(V(CFG))$, and
$\dots$ denotes the context and process id.

\textit{\textbf{Cell-by-cell translation}}: \label{tr_cell_by_cell}
The simplest translation in both implementation and formalization
is to use a key value pair predicate $C$ to encode
the cells.
$C$ stores the cell name as
the first argument, and the associated memory content as the second argument.
Here $\_$ denotes a placeholder.

\begin{center}
\begin{tabular}{LLL}
  & l_0 \inbrac a_0 \outbrac r_0, &
  C('a', \_), \dots\\
  C('a', \_), \dots, & l_1 \inbrac a_1 \outbrac r_1, &
  C('b', \_), \dots\\
  C('c', \_), \dots & l_2 \inbrac a_2 \outbrac r_2, &
  C('d', \_), \dots\\
  C('c', \_), \dots & l_3 \inbrac a_3 \outbrac r_3, &
  C('e', \_), \dots\\
  C('a', \_), \dots & l_4 \inbrac a_4 \outbrac r_4, &
  C('a', \_), \dots\\
  C('a', \_), \dots & l_5 \inbrac a_5 \outbrac r_5, &
  C('c', \_), \dots\\
  C('c', \_), \dots & l_6 \inbrac a_6 \outbrac r_6
\end{tabular}
\end{center}

A problem with this translation is
that a read operation often yields the same fact
on both the left and right side of an MSR rule, which forms a loop.
This style consistently leads to
the worst performance in most cases (see Figures~\ref{table:emverify-mastercard-cda-nopin-low}
and \ref{table:emverify-visa-dda-low}).

\textit{\textbf{Forward translation}}:\label{tr_forward}
In this translation, a rule sets up the context required by the successor rule
(we look ``forward'' so to speak).
The translation of Figure~\ref{fig:basic_cfg} then reads:

\begin{center}
\begin{tabular}{LLLL}
	0. & &l_0 \inbrac a_0 \outbrac r_0, & St(1, \dots)\\
	1_2. & St(1, \dots), & l_1 \inbrac a_1 \outbrac r_1, & St(2, \dots)\\
	1_3. & St(1, \dots), & l_1 \inbrac a_1 \outbrac r_1, & St(3, \dots)\\
	1_4. & St(1, \dots), & l_1 \inbrac a_1 \outbrac r_1, & St(4, \dots)\\
	2. & St(2, \dots), & l_2 \inbrac a_2 \outbrac r_2, & St(5, \dots)\\
	3. & St(3, \dots), & l_3 \inbrac a_3 \outbrac r_3, & St(5, \dots)\\
	4. & St(4, \dots), & l_4 \inbrac a_4 \outbrac r_4, & St(5, \dots)\\
	5. & St(5, \dots), & l_5 \inbrac a_5 \outbrac r_5, & St(6, \dots)\\
  6. & St(6, \dots), & l_6 \inbrac a_6 \outbrac r_6
\end{tabular}
\end{center}
We note that in this style,
joins are efficient (only one copy of rule $5$ is needed),
but splits can require potentially many duplications of rules
(rule $1$ requires 3 copies in this case to account for different
possible successors).

\textit{\textbf{Backward translation}}:\label{tr_backward}
In this style, a rule accesses the context left behind
by the predecessor rule (we look ``backward'').
The translation of Figure~\ref{fig:basic_cfg} then reads:

\begin{center}
\begin{tabular}{LLLL}
	0. & &l_0 \inbrac a_0 \outbrac r_0, & St(0, \dots)\\
	1. & St(0, \dots), & l_1 \inbrac a_1 \outbrac r_1, & St(1, \dots)\\
	2. & St(1, \dots), & l_2 \inbrac a_2 \outbrac r_2, & St(2, \dots)\\
	3. & St(1, \dots), & l_3 \inbrac a_3 \outbrac r_3, & St(3, \dots)\\
	4. & St(1, \dots), & l_4 \inbrac a_4 \outbrac r_4, & St(4, \dots)\\
	5_2. & St(2, \dots), & l_5 \inbrac a_5 \outbrac r_5, & St(5, \dots)\\
	5_3. & St(3, \dots), & l_5 \inbrac a_5 \outbrac r_5, & St(5, \dots)\\
	5_4. & St(4, \dots), & l_5 \inbrac a_5 \outbrac r_5, & St(5, \dots)\\
	6. & St(5, \dots), & l_6 \inbrac a_6 \outbrac r_6, & St(6, \dots)
\end{tabular}
\end{center}

Here we see opposite properties compared to forward translation:
namely splits are now efficient (only one copy of rule $1$ is needed),
but joins are inefficient (rule $5$ requires 3 copies
to account for different possible predecessors).

\subsection{Hybrid translation} \label{tr_hybrid}

In the simple case, it seems that we can use the forward translation
for splits and a backward translation for joins to mitigate all
shortcomings.
Indeed this is the usual natural manual encoding
style we see in case studies.
However, to accommodate for the general case,
we require a formalization of the intuition over
graphs (our intermediate representation)
rather than just over explicit splits and joins,
since we want the higher level construct
of Tamgram to remain flexible (should we introduce other
loop primitives, for instance).
We specify the translation from a Tamgram process (in CFG representation)
to Tamarin rules as follows, given a graph $\langle V, E \rangle$.
We denote the translation function of a Tamgram process as $T$,
which consists of a set of Tamarin multiset rewriting rules.

\begin{definition}
  $flatten(ctx)$ yields
  a tuple $\langle c_0, c_1, \dots, c_n \rangle$
  sorted lexicographically by the name of cells
  such that $\{ c_0, c_1, \dots, c_n \} = ctx$.
  Meaningfully we just need
  a deterministic mapping from a set of cells
  to a tuple of cells.
\end{definition}

\begin{definition}
  We define $cvar$ as follows:
  \begin{align*}
    cvar(c) &= \text{a unique name mapped to $c$}\\
    cvar(\langle x_0, \dots, x_n \rangle ) &= \langle cvar(x_0), \dots, cvar(x_n) \rangle\\
    cvar([ x_0, \dots, x_n ]) &= [ cvar(x_0), \dots, cvar(x_n) ]\\
    cvar(l \inbrac a \outbrac r) &= cvar(l) \inbrac cvar(a) \outbrac cvar(r)
  \end{align*}
  $cvar$ is essentially a function that replaces all cells with variables.
  This is largely a formality, and in implementation we can just create
  the mapping required by applying a reserved prefix to cell names.
  For example, cell $c$ can be replaced with variable $tgc\_c$ where $tgc\_$
  is the reserved prefix.
\end{definition}

Here we give the basic version of the definitions
of $ctx_{maxR}(k)$ and $ctx_{maxRA}(k)$.
Details are given in the technical report.

\begin{definition}
    We define $ctx_{maxR}(k)$ to be the ``maximal context required''
    by rule $k$, which is an over-approximation of cells that the process should carry whenever rule $k$ is reached to not
    result in access of undefined cells.
\end{definition}

\begin{definition}
    We define $ctx_{maxRA}(k)$ to be the ``maximal context required afterwards''
    by rule $k$, which is the union of $ctx_{maxR}$ of all
    successors $k'_0, k'_1, \dots$ of $k$.
    
    This may not always relate to $ctx_{maxR}(k)$
    by the subset relation. For example,
    if $k$ does not use cell $a$, but defines it
    on the right field,
    $a \notin ctx_{maxR}(k)$ but
    $a$ may still be in $ctx_{maxR}(k')$ for
    some successor $k'$.
\end{definition}

We annotated whether a rule should use forward or backward
style based on a simple heuristic called {\em exit bias}.
\begin{definition} \label{def:exit-bias}
  We define exit bias as follows: for any rule $ru$ indexed by $k$,\\
  \textbf{if} $succ(E, k) \leq 1$, \textbf{then} $exit\_bias(\langle V, E \rangle, k) = Forward$
  \\
  \textbf{else} $exit\_bias(\langle V, E \rangle, k) = Backward$
\end{definition}

For instance, in Figure~\ref{fig:basic_cfg},
rule 1 has a $Backward$ exit bias,
while all other rules have a $Forward$ exit bias.

We do not assign ``entry bias,''
as unifying entry biases and exit biases
introduces significant overhead without obvious benefits,
i.e.~the strictly ``optimal'' arrangement
does not seem to be significantly better for
the usual CFGs in our tests.
Instead, we create as many copies
as needed for a given rule
to match the exit biases of its
predecessors.

From the biases, we compute the
possible entry and exit facts, which
we use to compute the exact copies of the rules
to generate.

The set of possible exit facts
used by a rule is determined
by the exit bias. If the rule is
forward exit biased, then the rule
handles the context required for the successor (if exists).
Otherwise, the rule leaves a context
that satisfies the memory
requirement of all successors.

\begin{definition}
  We define the set of possible exit facts of a rule $ru$ indexed by $k$,\\
  \textbf{if} $exit\_bias(\langle V, E \rangle, k) = Forward$,
  \textbf{then} $\forall k' \in succ(E, k)$.
  \begin{align*}
    St^F (\sim pid, k', cvar(ctx_{maxR}(k'))) \in exit\_facts(\langle V, E \rangle, k)
  \end{align*}
  \textbf{else}
  \begin{align*}
      St^B(\sim pid, k, cvar(ctx_{maxRA}(k))) \in exit\_facts(\langle V, E \rangle, k)
  \end{align*}
\end{definition}

\begin{definition}
  We define the set of possible entry facts of a rule $ru$ indexed by $k$:
  $\forall k' \in pred(E, k)$,\\
  \textbf{if} $exit\_bias(\langle V, E \rangle, k') = Forward$,
  \textbf{then}
  \begin{align*}
      St^F(\sim pid, k, cvar(ctx_{maxR}(k))) \in entry\_facts(\langle V, E \rangle, k)
  \end{align*}
  \textbf{else}
  \begin{align*}
      St^B(\sim pid, k', cvar(ctx_{maxRA}(k'))) \in entry\_facts(\langle V, E \rangle, k)
  \end{align*}
which essentially reads we accommodate the exit bias of each predecessor.
\end{definition}

\begin{definition}
  For a Tamgram system $Sys$, we define the translation
  of the system into set of Tamarin rules $T(Sys)$ as
  follows: $\forall$ rule $l \inbrac a \outbrac r$ indexed by $k$
  \text{ . }
  let $entryFs = entry\_facts(\langle V, E \rangle, k)$
  and $exitFs = exit\_facts(\langle V, E \rangle, k)$
  \textbf{if} $entryFs = \emptyset$, \textbf{then}
  $\forall exitF \in exitFs$.
  \begin{align*}
    cvar(l \inbrac a \outbrac [exitF, r]) \in T(Sys)
  \end{align*}
  \textbf{else if} $exitFs = \emptyset$, \textbf{then}
  $\forall entryF \in entryFs$.
  \begin{align*}
    cvar([entryF, l] \inbrac a \outbrac r) \in T(Sys)
  \end{align*}
  \textbf{else}
  $\forall entryF \in entryFs, exitF \in exitFs$.
  \begin{align*}
    cvar([entryF, l] \inbrac a \outbrac [exitF, r]) \in T(Sys)
  \end{align*}
\end{definition}

We demonstrate this in action by using the basic CFG above.
We list the exit biases, possible exit facts of all the rules in accordance
with above definitions.
\begin{center}
\begin{supertabular}{LLLL}
	   & \text{Exit bias} & \text{Exit facts} & \text{Entry facts} \\
	0. & Forward & St^F(\sim pid, 1, \dots) & \\
    1. & Backward & St^B(\sim pid, 1, \dots) & St^F(\sim pid, 1, \dots) \\
	2. & Forward & St^F(\sim pid, 5, \dots) & St^B(\sim pid, 1, \dots) \\
	3. & Forward & St^F(\sim pid, 5, \dots) & St^B(\sim pid, 1, \dots) \\
	4. & Forward & St^F(\sim pid, 5, \dots) & St^B(\sim pid, 1, \dots) \\
	5  & Forward & St^F(\sim pid, 6, \dots) & St^F(\sim pid, 5, \dots) \\
	6. & Forward &  & St^F(\sim pid, 6, \dots)\\
\end{supertabular}
\end{center}
In this case, as each rule has at most one possible exit fact
and at most one possible entry fact,
we only need to generate one copy of each rule:
\begin{center}
\begin{supertabular}{LLLL}
	0. & &l_0 \inbrac a_0 \outbrac r_0, & St^F(\sim pid, 1, \dots)\\
	1. & St^F(\sim pid, 1, \dots), & l_1 \inbrac a_1 \outbrac r_1, & St^B(\sim pid, 1, \dots)\\
	2. & St^B(\sim pid, 1, \dots), & l_2 \inbrac a_2 \outbrac r_2, & St^F(\sim pid, 5, \dots)\\
	3. & St^B(\sim pid, 1, \dots), & l_3 \inbrac a_3 \outbrac r_3, & St^F(\sim pid, 5, \dots)\\
	4. & St^B(\sim pid, 1, \dots), & l_4 \inbrac a_4 \outbrac r_4, & St^F(\sim pid, 5, \dots)\\
	5. & St^F(\sim pid, 5, \dots), & l_5 \inbrac a_5 \outbrac r_5, & St^F(\sim pid, 6, \dots)\\
	6. & St^F(\sim pid, 6, \dots), & l_6 \inbrac a_6 \outbrac r_6 & \\
\end{supertabular}
\end{center}

Overall, this style combines the strength of forward and backward
translation styles with relatively
straightforward procedures.

\section{Experiments on existing case studies}

To demonstrate and test the practicality of our
procedures, we performed a case study on one set of small protocols (CSF18 XOR),
and one large protocol with different variants (EMVerify Contactless).
Our primary concern lies in the latter.

All tests are done with the following parameters passed
to Haskell runtime when invoking Tamarin:
4 CPU cores (2.1GHz), 50GB max memory.
Tamarin is terminated at 60 minutes mark if not finished,
(denoted by \texttt{TIMEOUT} in tables).

We tried to be as faithful to the original Tamarin specification
files as we could during our translation,
with the only alteration being conversion of state predicates
to uses of cells, and replicating the control flow
as denoted by the original state predicates.
But we cannot rule out discrepancies in our translation,
as shown by some rows where some compiled model
is vastly slower or faster than the original version,
even though it is comparable in most other rows.

For CSF18 XOR, as the cases do not share the same
set of lemmas, we only present a summary in
Figure~\ref{table:csf18-summary}
(see Appendix~\ref{appendix:csf18-full} for details).
For EMVerify, due to the large number of variants,
and significantly longer run time
for each lemma,
we only picked two variants with the fewest
timeout cases to illustrate for all
translation styles,
but we note the performance difference is usually
consistent across variants in our testing
(see Appendix~\ref{appendix:emverify-full} for details).

\subsection{CSF18 XOR}

\begin{figure}[h]
    \fontsize{5}{6}
      \selectfont
      
            \begin{tblr}{
                    hlines,
                    vlines,
                    colspec={c 
        *{1}{p{0.9cm}}*{1}{p{0.9cm}}*{1}{p{0.9cm}}*{1}{p{0.9cm}}*{1}{p{0.9cm}}
                    },
                }
        & Original& Cell-by-cell& Forward& Backward& Hybrid\\
CH07& 2.0& 3.4& 2.2& 2.2& 2.1\\
CRxor& 1.2& 1.2& 1.2& 1.2& 1.2\\
KCL07& 1.3& 1.5& 1.4& 1.5& 1.4\\
LAK06& 9.8& 12.7& 11.5& 11.1& 11.1\\
NSLPK3xor& 3.3& 10.7& 3.9& 3.7& 3.9\\
\end{tblr}
      
      \caption{CSF18 XOR benchmark summary (seconds)}
      \label{table:csf18-summary}
\end{figure}

For CSF18 XOR \cite{csf18-xor},
all the protocols are relatively simple in structure,
with the CFG of a process mostly being a line,
i.e.~there is no (conditional) branching.

All translation styles
(Figure~\ref{table:csf18-ch07} to\ref{table:csf18-nslpk3xor}
in Appendix)
give comparable overheads for most of the cases,
and are within margin of error with
respect to the measurement of
the original specification.

\subsection{EMVerify Contactless}

\begin{figure}[h]
      \fontsize{5}{6}
      \selectfont

            \begin{tblr}{
                    hlines,
                    vlines,
                    colspec={c 
        *{1}{p{0.9cm}}*{1}{p{0.9cm}}*{1}{p{0.9cm}}*{1}{p{0.9cm}}*{1}{p{0.9cm}}
                    },
                }
        & Original& Cell-by-cell& Forward& Backward& Hybrid\\
executable& 55.5& \texttt{TIMEOUT}& 91.2& 78.3& 63.8\\
bank\_accepts& 707.7& \texttt{TIMEOUT}& 1018.9& 1006.7& 932.4\\
auth\_to\_terminal\_minimal& \texttt{TIMEOUT}& \texttt{TIMEOUT}& \texttt{TIMEOUT}& \texttt{TIMEOUT}& \texttt{TIMEOUT}\\
auth\_to\_terminal& \texttt{TIMEOUT}& \texttt{TIMEOUT}& \texttt{TIMEOUT}& \texttt{TIMEOUT}& \texttt{TIMEOUT}\\
auth\_to\_bank\_minimal& 60.9& \texttt{TIMEOUT}& 98.5& 89.7& 66.9\\
auth\_to\_bank& 619.0& \texttt{TIMEOUT}& 206.8& 203.4& 172.1\\
secrecy\_MK& 46.9& \texttt{TIMEOUT}& 77.6& 62.5& 49.2\\
secrecy\_privkCard& 45.1& \texttt{TIMEOUT}& 81.8& 74.7& 47.4\\
secrecy\_PAN& 46.0& \texttt{TIMEOUT}& 76.4& 68.0& 47.9\\
\end{tblr}
      
      \caption{EMVerify Contactless Mastercard CDA NoPIN Low benchmark (seconds)}
      \label{table:emverify-mastercard-cda-nopin-low}
\end{figure}

\begin{figure}[h]
      \fontsize{5}{6}
      \selectfont

            \begin{tblr}{
                    hlines,
                    vlines,
                    colspec={c 
        *{1}{p{0.9cm}}*{1}{p{0.9cm}}*{1}{p{0.9cm}}*{1}{p{0.9cm}}*{1}{p{0.9cm}}
                    },
                }
        & Original& Cell-by-cell& Forward& Backward& Hybrid\\
executable& 54.6& \texttt{TIMEOUT}& 91.3& 77.3& 59.4\\
bank\_accepts& 1261.0& \texttt{TIMEOUT}& 951.6& 933.5& 872.1\\
auth\_to\_terminal\_minimal& \texttt{TIMEOUT}& \texttt{TIMEOUT}& 978.2& 972.4& 923.4\\
auth\_to\_bank\_minimal& 82.4& \texttt{TIMEOUT}& 154.2& 103.5& 91.7\\
auth\_to\_bank& 309.3& \texttt{TIMEOUT}& 672.1& 383.8& 334.2\\
secrecy\_MK& 47.2& \texttt{TIMEOUT}& 71.8& 62.4& 47.8\\
secrecy\_privkCard& 47.7& \texttt{TIMEOUT}& 73.6& 69.8& 48.4\\
secrecy\_PIN& 44.4& \texttt{TIMEOUT}& 83.1& 76.5& 47.0\\
secrecy\_PAN& 43.9& \texttt{TIMEOUT}& 80.8& 69.2& 51.2\\
\end{tblr}
      
      \caption{EMVerify Contactless Visa DDA Low benchmark (seconds)}
      \label{table:emverify-visa-dda-low}
\end{figure}

For EMVerify \cite{emverify} cases,
the translation style has a much higher impact
in the verification time.

\begin{itemize}
  \item
  The Cell-by-cell translation, which
  is the simplest,
  consistently times out 
  at 1 hour mark for all
  lemmas.
  
  \item
  Both the forward and backward translations
  finish within a reasonable time frame
  w.r.t. the original model,
  but consistently carry a 50\% to 85\%
  verification time overhead.
  
  \item
  The hybrid translation closely matches
  the original model in verification time
  in the majority of the cases.
\end{itemize}

\section{Conclusions and future work}

Tamgram 
offers some of the modern programming
experiences
to Tamarin users
while preserving the strength
of Tamarin.
We believe that the addition
of features is necessary
for writing readable
models of complex security protocols.

We identify several avenues for further
improvements:

\begin{itemize}
  \item
  We intend to further investigate
  whether the clearer specification
  of the control flow of protocol in the form
  of Tamgram processes allows for
  better automated derivation of proof
  heuristics (in the form of Tamarin oracles).
  High-level languages such as SAPIC
  use a custom proof oracle
  to guide Tamarin during verification
  of the compiled model.
  The custom proof heuristics
  lead to siginificant speedup in
  some cases \cite{SmartVerif}.

  Approaches based on dynamic proof heuristics
  such as SmartVerif \cite{SmartVerif}
  also
  further improve
  the degree of automation.
  
  \item
  Other projects, such as
  SAPIC+ \cite{sapic-plus}, carry an
  optimization pass during translation into
  Tamarin.
  This yields significant speedup
  as a result in some protocols.
  These are possible due to the protocol
  being specified in a higher level language
  where intentions are more clearly captured.

  We intend to investigate investigate if Tamgram
  may benefit from similar optimization
  procedures.

  \item
  In very complex case studies,
  such as the WPA2 case study, several advanced features
  related to the tuning of heuristics for Tamarin
  resolution are used. Tamgram currently lacks
  the syntactic support for these.
  Furthermore, it remains to be
  investigated how we should handle
  these heuristics hints during
  the various translation passes.

  \item
  Currently, it can be difficult
  to interpret Tamarin proofs
  based on the compiled results because
  there is a disconnect
  between the compiled model and
  the Tamgram model.
  We have begun work
  on the idea of ``back translation'',
  to rewrite the proof graph
  in Tamarin interactive mode
  to map closely
  to the original Tamgram model instead
  of the compiled model.
\end{itemize}

\clearpage

\bibliographystyle{IEEEtran}
\bibliography{refs}

\clearpage
\begin{appendices}

\onecolumn

\section{CSF18 XOR full benchmark} \label{appendix:csf18-full}

\begin{figure*}[h]
  \begin{minipage}{\linewidth}
    \begin{minipage}{0.5\linewidth}
      \fontsize{5}{6}
      \selectfont

            \begin{tblr}{
                    hlines,
                    vlines,
                    colspec={c 
        *{1}{p{0.75cm}}*{1}{p{0.75cm}}*{1}{p{0.75cm}}*{1}{p{0.75cm}}*{1}{p{0.75cm}}
                    },
                }
        & Original& Cell-by-cell& Forward& Backward& Hybrid\\
recentalive\_tag& 0.6& 0.7& 0.7& 0.7& 0.7\\
recentalive\_reader& 0.3& 0.7& 0.4& 0.4& 0.4\\
noninjectiveagreement\_tag& 0.3& 0.8& 0.4& 0.4& 0.3\\
noninjectiveagreement\_reader& 0.3& 0.4& 0.3& 0.3& 0.3\\
executable& 0.4& 0.7& 0.4& 0.4& 0.5\\
\end{tblr}
      
      \caption{CSF18 XOR CH07}
      \label{table:csf18-ch07}
    \end{minipage}
    \begin{minipage}{0.5\linewidth}
      \fontsize{5}{6}
      \selectfont

            \begin{tblr}{
                    hlines,
                    vlines,
                    colspec={c 
        *{1}{p{0.75cm}}*{1}{p{0.75cm}}*{1}{p{0.75cm}}*{1}{p{0.75cm}}*{1}{p{0.75cm}}
                    },
                }
        & Original& Cell-by-cell& Forward& Backward& Hybrid\\
alive& 0.3& 0.3& 0.3& 0.3& 0.3\\
recentalive\_tag& 0.4& 0.5& 0.5& 0.5& 0.5\\
executable& 0.4& 0.4& 0.4& 0.4& 0.4\\
\end{tblr}
      
      \caption{CSF18 XOR CRxor}
      \label{table:csf18-crxor}
    \end{minipage}
  \end{minipage}
\end{figure*}

\begin{figure*}[h]
  \begin{minipage}{\linewidth}
    \begin{minipage}{0.5\linewidth}
      \fontsize{5}{6}
      \selectfont

            \begin{tblr}{
                    hlines,
                    vlines,
                    colspec={c 
        *{1}{p{0.75cm}}*{1}{p{0.75cm}}*{1}{p{0.75cm}}*{1}{p{0.75cm}}*{1}{p{0.75cm}}
                    },
                }
        & Original& Cell-by-cell& Forward& Backward& Hybrid\\
recentalive\_tag& 0.9& 1.1& 1.0& 1.0& 1.0\\
executable& 0.4& 0.4& 0.4& 0.4& 0.4\\
\end{tblr}
      
      \caption{CSF18 XOR KCL07}
      \label{table:csf18-kcl07}
    \end{minipage}
    \begin{minipage}{0.5\linewidth}
      \fontsize{5}{6}
      \selectfont

            \begin{tblr}{
                    hlines,
                    vlines,
                    colspec={c 
        *{1}{p{0.75cm}}*{1}{p{0.75cm}}*{1}{p{0.75cm}}*{1}{p{0.75cm}}*{1}{p{0.75cm}}
                    },
                }
        & Original& Cell-by-cell& Forward& Backward& Hybrid\\
executable& 0.8& 0.6& 0.9& 0.8& 0.9\\
helpingSecrecy& 0.2& 0.3& 0.2& 0.2& 0.2\\
noninjectiveagreementTAG& 8.2& 11.2& 9.6& 9.3& 9.2\\
noninjectiveagreementREADER& 0.6& 0.5& 0.8& 0.8& 0.8\\
\end{tblr}
      
      \caption{CSF18 XOR LAK06}
      \label{table:csf18-lak06}
    \end{minipage}
  \end{minipage}
\end{figure*}

\begin{figure*}[h]
  \begin{minipage}{\linewidth}
    \begin{minipage}{0.5\linewidth}
      \fontsize{5}{6}
      \selectfont

            \begin{tblr}{
                    hlines,
                    vlines,
                    colspec={c 
        *{1}{p{0.75cm}}*{1}{p{0.75cm}}*{1}{p{0.75cm}}*{1}{p{0.75cm}}*{1}{p{0.75cm}}
                    },
                }
        & Original& Cell-by-cell& Forward& Backward& Hybrid\\
types& 0.5& 1.2& 0.6& 0.5& 0.6\\
nonce\_secrecy& 1.2& 2.1& 1.5& 1.3& 1.4\\
injective\_agree& 1.2& 6.3& 1.3& 1.4& 1.4\\
session\_key\_setup\_possible& 0.4& 1.1& 0.5& 0.5& 0.4\\
\end{tblr}
      
      \caption{CSF18 XOR NSLPK3xor}
      \label{table:csf18-nslpk3xor}
    \end{minipage}
  \end{minipage}
\end{figure*}

\section{EMVerify hybrid translation benchmark} \label{appendix:emverify-full}

\begin{figure}[h]
  \fontsize{5}{6}
  \selectfont
  Legend:
  \begin{tblr}{
      hlines,
      vlines,
      colspec={*{2}{p{2cm}}},
      column{2}={blue8},
    }
    Original & Hybrid
  \end{tblr}

            \begin{tblr}{
                    hlines,
                    vlines,
                    colspec={c 
        *{1}{p{1.5cm}} *{1}{p{1cm}}*{1}{p{1.5cm}} *{1}{p{1cm}}*{1}{p{1.5cm}} *{1}{p{1cm}}*{1}{p{1.5cm}} *{1}{p{1cm}}
                    },
                    column{4-5,8-9}={blue8},
                }
        & \SetCell[c=4]{} executable & & && \SetCell[c=4]{} bank\_accepts & & &\\
Mastercard\_CDA\_NoPIN\_High& falsified 47 steps& 48.2& falsified 48 steps& 51.1& $\times$& -& $\times$& -\\
Mastercard\_CDA\_NoPIN\_Low& verified 30 steps& 55.5& verified 28 steps& 63.8& verified 2897 steps& 707.7& verified 3629 steps& 932.4\\
Mastercard\_CDA\_OnlinePIN\_High& verified 30 steps& 55.6& verified 29 steps& 69.4& verified 7145 steps& 1715.6& verified 9331 steps& 2396.0\\
Mastercard\_CDA\_OnlinePIN\_Low& verified 30 steps& 55.7& verified 28 steps& 64.2& verified 2905 steps& 707.2& verified 3637 steps& 973.5\\
Mastercard\_DDA\_NoPIN\_High& falsified 49 steps& 46.8& falsified 50 steps& 52.1& $\times$& -& $\times$& -\\
Mastercard\_DDA\_NoPIN\_Low& verified 29 steps& 54.1& verified 27 steps& 56.2& falsified 16 steps& 344.9& falsified 17 steps& 603.0\\
Mastercard\_DDA\_OnlinePIN\_High& verified 29 steps& 54.1& verified 28 steps& 58.6& verified 4865 steps& 1054.8& verified 6371 steps& 1468.1\\
Mastercard\_DDA\_OnlinePIN\_Low& verified 29 steps& 51.3& verified 27 steps& 57.4& falsified 16 steps& 346.5& falsified 17 steps& 594.9\\
Mastercard\_SDA\_NoPIN\_High& falsified 45 steps& 47.4& falsified 46 steps& 49.9& $\times$& -& $\times$& -\\
Mastercard\_SDA\_NoPIN\_Low& verified 27 steps& 51.1& verified 25 steps& 54.3& falsified 13 steps& 189.8& falsified 14 steps& 52.7\\
Mastercard\_SDA\_OnlinePIN\_High& verified 27 steps& 51.1& verified 26 steps& 58.8& verified 2315 steps& 515.9& verified 3047 steps& 707.9\\
Mastercard\_SDA\_OnlinePIN\_Low& verified 27 steps& 51.0& verified 25 steps& 53.1& falsified 13 steps& 192.4& falsified 14 steps& 53.5\\
Visa\_DDA\_High& verified 29 steps& 53.3& verified 29 steps& 60.2& verified 5617 steps& 933.8& verified 5657 steps& 985.7\\
Visa\_DDA\_Low& verified 28 steps& 54.6& verified 28 steps& 59.4& falsified 17 steps& 1261.0& falsified 18 steps& 872.1\\
Visa\_DDA\_Low\_Fix& verified 28 steps& 63.3& verified 28 steps& 71.1& verified 3710 steps& 733.8& verified 3728 steps& 779.6\\
Visa\_EMV\_High& verified 22 steps& 49.0& verified 23 steps& 53.1& verified 308 steps& 69.3& verified 356 steps& 74.8\\
Visa\_EMV\_Low& verified 22 steps& 49.2& verified 23 steps& 51.2& verified 210 steps& 61.9& verified 242 steps& 70.6\\
\end{tblr}

  \caption{EMVerify table 1}
  \label{table:emverify1}
\end{figure}

\begin{figure}[h]
  \fontsize{5}{6}
  \selectfont

            \begin{tblr}{
                    hlines,
                    vlines,
                    colspec={c 
        *{1}{p{1.5cm}} *{1}{p{1cm}}*{1}{p{1.5cm}} *{1}{p{1cm}}*{1}{p{1.5cm}} *{1}{p{1cm}}*{1}{p{1.5cm}} *{1}{p{1cm}}
                    },
                    column{4-5,8-9}={blue8},
                }
        & \SetCell[c=4]{} auth\_to\_terminal\_minimal & & && \SetCell[c=4]{} auth\_to\_terminal & & &\\
Mastercard\_CDA\_NoPIN\_High& $\times$& -& $\times$& -& $\times$& -& $\times$& -\\
Mastercard\_CDA\_NoPIN\_Low& $\times$& \texttt{TIMEOUT}& $\times$& \texttt{TIMEOUT}& $\times$& \texttt{TIMEOUT}& $\times$& \texttt{TIMEOUT}\\
Mastercard\_CDA\_OnlinePIN\_High& $\times$& \texttt{TIMEOUT}& $\times$& \texttt{TIMEOUT}& $\times$& \texttt{TIMEOUT}& $\times$& \texttt{TIMEOUT}\\
Mastercard\_CDA\_OnlinePIN\_Low& $\times$& \texttt{TIMEOUT}& $\times$& \texttt{TIMEOUT}& $\times$& \texttt{TIMEOUT}& $\times$& \texttt{TIMEOUT}\\
Mastercard\_DDA\_NoPIN\_High& $\times$& -& $\times$& -& $\times$& -& $\times$& -\\
Mastercard\_DDA\_NoPIN\_Low& $\times$& \texttt{TIMEOUT}& falsified 13 steps& 246.1& $\times$& -& $\times$& -\\
Mastercard\_DDA\_OnlinePIN\_High& $\times$& \texttt{TIMEOUT}& $\times$& \texttt{TIMEOUT}& $\times$& \texttt{TIMEOUT}& $\times$& \texttt{TIMEOUT}\\
Mastercard\_DDA\_OnlinePIN\_Low& $\times$& \texttt{TIMEOUT}& falsified 13 steps& 250.0& $\times$& -& $\times$& -\\
Mastercard\_SDA\_NoPIN\_High& $\times$& -& $\times$& -& $\times$& -& $\times$& -\\
Mastercard\_SDA\_NoPIN\_Low& $\times$& \texttt{TIMEOUT}& falsified 10 steps& 536.6& $\times$& -& $\times$& -\\
Mastercard\_SDA\_OnlinePIN\_High& $\times$& \texttt{TIMEOUT}& $\times$& \texttt{TIMEOUT}& $\times$& \texttt{TIMEOUT}& $\times$& \texttt{TIMEOUT}\\
Mastercard\_SDA\_OnlinePIN\_Low& $\times$& \texttt{TIMEOUT}& falsified 10 steps& 542.0& $\times$& -& $\times$& -\\
Visa\_DDA\_High& $\times$& \texttt{TIMEOUT}& $\times$& \texttt{TIMEOUT}& $\times$& \texttt{TIMEOUT}& $\times$& \texttt{TIMEOUT}\\
Visa\_DDA\_Low& $\times$& \texttt{TIMEOUT}& falsified 14 steps& 923.4& $\times$& -& $\times$& -\\
Visa\_DDA\_Low\_Fix& $\times$& \texttt{TIMEOUT}& $\times$& \texttt{TIMEOUT}& $\times$& \texttt{TIMEOUT}& $\times$& \texttt{TIMEOUT}\\
Visa\_EMV\_High& falsified 11 steps& 74.7& falsified 11 steps& 85.2& $\times$& -& $\times$& -\\
Visa\_EMV\_Low& falsified 11 steps& 75.7& falsified 11 steps& 80.0& $\times$& -& $\times$& -\\
\end{tblr}
  
  \caption{EMVerify table 2}
  \label{table:emverify2}
\end{figure}

\begin{figure}[h]
  \fontsize{5}{6}
  \selectfont

            \begin{tblr}{
                    hlines,
                    vlines,
                    colspec={c 
        *{1}{p{1.5cm}} *{1}{p{1cm}}*{1}{p{1.5cm}} *{1}{p{1cm}}*{1}{p{1.5cm}} *{1}{p{1cm}}*{1}{p{1.5cm}} *{1}{p{1cm}}
                    },
                    column{4-5,8-9}={blue8},
                }
        & \SetCell[c=4]{} auth\_to\_bank\_minimal & & && \SetCell[c=4]{} auth\_to\_bank & & &\\
Mastercard\_CDA\_NoPIN\_High& $\times$& -& $\times$& -& $\times$& -& $\times$& -\\
Mastercard\_CDA\_NoPIN\_Low& verified 141 steps& 60.9& verified 147 steps& 66.9& verified 2818 steps& 619.0& verified 612 steps& 172.1\\
Mastercard\_CDA\_OnlinePIN\_High& verified 221 steps& 74.9& verified 260 steps& 86.5& verified 3566 steps& 820.6& verified 1086 steps& 293.0\\
Mastercard\_CDA\_OnlinePIN\_Low& verified 141 steps& 63.6& verified 147 steps& 69.6& verified 2818 steps& 616.9& verified 612 steps& 176.3\\
Mastercard\_DDA\_NoPIN\_High& $\times$& -& $\times$& -& $\times$& -& $\times$& -\\
Mastercard\_DDA\_NoPIN\_Low& verified 137 steps& 59.5& verified 143 steps& 68.7& verified 2912 steps& 490.3& verified 666 steps& 157.5\\
Mastercard\_DDA\_OnlinePIN\_High& verified 203 steps& 72.5& verified 242 steps& 81.6& verified 3372 steps& 621.6& verified 1116 steps& 264.0\\
Mastercard\_DDA\_OnlinePIN\_Low& verified 137 steps& 61.6& verified 143 steps& 65.6& verified 2912 steps& 497.8& verified 666 steps& 155.4\\
Mastercard\_SDA\_NoPIN\_High& $\times$& -& $\times$& -& $\times$& -& $\times$& -\\
Mastercard\_SDA\_NoPIN\_Low& verified 102 steps& 56.2& verified 108 steps& 61.7& verified 1377 steps& 284.5& verified 363 steps& 101.1\\
Mastercard\_SDA\_OnlinePIN\_High& verified 150 steps& 64.9& verified 183 steps& 71.6& verified 1688 steps& 355.2& verified 585 steps& 150.5\\
Mastercard\_SDA\_OnlinePIN\_Low& verified 102 steps& 57.8& verified 108 steps& 57.3& verified 1377 steps& 287.8& verified 363 steps& 100.2\\
Visa\_DDA\_High& verified 205 steps& 70.6& verified 205 steps& 74.7& verified 953 steps& 168.7& verified 953 steps& 182.2\\
Visa\_DDA\_Low& verified 337 steps& 82.4& verified 337 steps& 91.7& verified 1922 steps& 309.3& verified 1922 steps& 334.2\\
Visa\_DDA\_Low\_Fix& verified 310 steps& 96.8& verified 310 steps& 99.9& verified 1382 steps& 283.2& verified 1382 steps& 308.8\\
Visa\_EMV\_High& falsified 11 steps& 55.2& falsified 11 steps& 55.1& $\times$& -& $\times$& -\\
Visa\_EMV\_Low& falsified 11 steps& 62.7& falsified 11 steps& 65.3& $\times$& -& $\times$& -\\
\end{tblr}
  
  \caption{EMVerify table 3}
  \label{table:emverify3}
\end{figure}

\begin{figure}[h]
  \fontsize{5}{6}
  \selectfont

            \begin{tblr}{
                    hlines,
                    vlines,
                    colspec={c 
        *{1}{p{1.5cm}} *{1}{p{1cm}}*{1}{p{1.5cm}} *{1}{p{1cm}}*{1}{p{1.5cm}} *{1}{p{1cm}}*{1}{p{1.5cm}} *{1}{p{1cm}}
                    },
                    column{4-5,8-9}={blue8},
                }
        & \SetCell[c=4]{} secrecy\_MK & & && \SetCell[c=4]{} secrecy\_privkCard & & &\\
Mastercard\_CDA\_NoPIN\_High& verified 12 steps& 44.9& verified 12 steps& 50.8& verified 13 steps& 44.8& verified 14 steps& 47.3\\
Mastercard\_CDA\_NoPIN\_Low& verified 12 steps& 46.9& verified 12 steps& 49.2& verified 13 steps& 45.1& verified 14 steps& 47.4\\
Mastercard\_CDA\_OnlinePIN\_High& verified 12 steps& 43.3& verified 12 steps& 51.3& verified 13 steps& 46.5& verified 14 steps& 48.1\\
Mastercard\_CDA\_OnlinePIN\_Low& verified 12 steps& 44.9& verified 12 steps& 51.7& verified 13 steps& 46.2& verified 14 steps& 49.3\\
Mastercard\_DDA\_NoPIN\_High& verified 12 steps& 47.7& verified 12 steps& 51.3& verified 13 steps& 45.5& verified 14 steps& 51.8\\
Mastercard\_DDA\_NoPIN\_Low& verified 12 steps& 47.1& verified 12 steps& 46.7& verified 13 steps& 46.2& verified 14 steps& 46.8\\
Mastercard\_DDA\_OnlinePIN\_High& verified 12 steps& 44.9& verified 12 steps& 49.0& verified 13 steps& 46.4& verified 14 steps& 50.4\\
Mastercard\_DDA\_OnlinePIN\_Low& verified 12 steps& 44.4& verified 12 steps& 52.1& verified 13 steps& 47.3& verified 14 steps& 47.1\\
Mastercard\_SDA\_NoPIN\_High& verified 12 steps& 47.6& verified 12 steps& 50.3& $\times$& -& $\times$& -\\
Mastercard\_SDA\_NoPIN\_Low& verified 12 steps& 44.4& verified 12 steps& 47.0& $\times$& -& $\times$& -\\
Mastercard\_SDA\_OnlinePIN\_High& verified 12 steps& 48.2& verified 12 steps& 51.9& $\times$& -& $\times$& -\\
Mastercard\_SDA\_OnlinePIN\_Low& verified 12 steps& 48.7& verified 12 steps& 48.3& $\times$& -& $\times$& -\\
Visa\_DDA\_High& verified 12 steps& 43.8& verified 12 steps& 48.3& verified 13 steps& 47.5& verified 14 steps& 50.1\\
Visa\_DDA\_Low& verified 12 steps& 47.2& verified 12 steps& 47.8& verified 13 steps& 47.7& verified 14 steps& 48.4\\
Visa\_DDA\_Low\_Fix& verified 12 steps& 55.0& verified 12 steps& 58.2& verified 13 steps& 59.3& verified 14 steps& 60.2\\
Visa\_EMV\_High& verified 12 steps& 46.4& verified 12 steps& 51.9& verified 13 steps& 44.1& verified 14 steps& 47.1\\
Visa\_EMV\_Low& verified 12 steps& 47.1& verified 12 steps& 51.8& verified 13 steps& 44.7& verified 14 steps& 51.6\\
\end{tblr}
  
  \caption{EMVerify table 4}
  \label{table:emverify4}
\end{figure}

\clearpage
\twocolumn

\section{Typing}

The typing rules are outlined below.
The purpose of typing phase is to make further
restrictions on structure of terms,
and to subsume most of Tamarin's wellformedness check
so errors are detected earlier rather than later in the workflow.
Note that $bitstring$ is shortened to $bits$.

\vspace{1em}

\begin{align*}
  \inferrule[]
  {\text{$s$ is a string literal}}
  {\Gamma \vdash s \Rightarrow bits}
\end{align*}

\begin{align*}
  \inferrule[]
  {\text{f is an uninterpreted function symbol of arity N}}
  {\Gamma \vdash f \Rightarrow bits \times bits \times \dots \rightarrow bits}
\end{align*}

\begin{align*}
  \inferrule[]
  {\text{f is an uninterpreted predicate symbol of arity N}}
  {\Gamma \vdash f \Rightarrow bits \times bits \times \dots \rightarrow fact}
\end{align*}

\begin{align*}
  \inferrule[]
  {\text{$f$ is an uninterpreted action predicate symbol of arity N}}
  {\Gamma \vdash f \Rightarrow bits \times bits \times \dots \rightarrow afact}
\end{align*}

\begin{align*}
  \inferrule[]
  {\Gamma \vdash e1 \Rightarrow t_{e1}\\
  \Gamma, \text{x} \Rightarrow t_{e1} \vdash e_2 \Rightarrow t_{e2}}
  {\Gamma \vdash \text{let $x : t_{e1} = e_1$ in $e2$ } \Rightarrow t_{e2}}
\end{align*}

\begin{align*}
  \inferrule[]
  {\Gamma, x_1 \Rightarrow t_1, \dots \vdash e_1 \Rightarrow t_{e1}\\
  \Gamma, \text{f} \Rightarrow t_1 \times \dots \rightarrow t_{e1}
  \vdash e_2 \Rightarrow t_{e2}}
  {\Gamma \vdash \text{let $f(x_1, \dots) : t_{e1} = e_1$ in $e2$ } \Rightarrow t_{e2}}
\end{align*}

\begin{align*}
  \inferrule[]
  {\Gamma \vdash \text{f} \Rightarrow t_1 \times \dots \rightarrow t\\
  \Gamma \vdash x_1 \Rightarrow t_1\\
  \dots
  }
  {\Gamma \vdash f(x_1, \dots) \Rightarrow t }
\end{align*}

\begin{align*}
  \inferrule[]
  {\text{}}
  {\Gamma \vdash {'c} \Rightarrow cell }
\end{align*}

\begin{align*}
  \inferrule[]
  {\Gamma \vdash e \Rightarrow t }
  {\Gamma \vdash {'c} := e \Rightarrow statement }
\end{align*}

\begin{align*}
  \inferrule[]
  {\Gamma \vdash e \Rightarrow t }
  {\Gamma \vdash \text{$'c$ cas e $\Rightarrow patmatch$ }}
\end{align*}

\begin{align*}
  \inferrule[]
  {\Gamma \vdash e \Rightarrow t }
  {\Gamma \vdash \text{$e$ as $x$ $\Rightarrow t$ }}
\end{align*}

\begin{align*}
  \inferrule[]
  {\Gamma \vdash e \Rightarrow patmatch }
  {\Gamma \vdash e \Rightarrow ruleL' }
\end{align*}

\begin{align*}
  \inferrule[]
  {\Gamma \vdash e \Rightarrow fact }
  {\Gamma \vdash e \Rightarrow ruleL'
  }
\end{align*}

\begin{align*}
\inferrule[]
{\Gamma \vdash e \Rightarrow afact }
{\Gamma \vdash e \Rightarrow ruleA' }
\end{align*}

\begin{align*}
  \inferrule[]
  {\Gamma \vdash e \Rightarrow fact }
  {\Gamma \vdash e \Rightarrow ruleR'
  }
\end{align*}

\begin{align*}
\inferrule[]
{\Gamma \vdash e \Rightarrow statement }
{\Gamma \vdash e \Rightarrow ruleR' }
\end{align*}

\begin{align*}
\inferrule[]
{\Gamma \vdash e_1 \Rightarrow ruleR'\\
 \dots
}
{\Gamma \vdash [e_1, \dots] \Rightarrow ruleR }
\end{align*}

\begin{align*}
  \inferrule[]
  { \Gamma \vdash r \Rightarrow ruleR
  }
  { \Gamma \vdash r \Rightarrow ruleAR }
\end{align*}

\begin{align*}
\inferrule[]
{ \Gamma \vdash e_1 \Rightarrow ruleA'\\
	\dots\\
	\Gamma \vdash r \Rightarrow ruleR
}
{\Gamma \vdash [e_1, \dots \outbrac r \Rightarrow ruleAR }
\end{align*}

\begin{align*}
\inferrule[]
{ \Gamma \vdash e_1 \Rightarrow ruleL'\\
	\dots\\
	\Gamma \vdash ar \Rightarrow ruleAR
}
{\Gamma \vdash [e_1, \dots \outbrac ar \Rightarrow rule }
\end{align*}

\begin{align*}
  \inferrule[]
  {\text{}}
  {\Gamma \vdash \mathbf{0} \Rightarrow process }
\end{align*}

\begin{align*}
  \inferrule[]
  {\Gamma \vdash ru \Rightarrow rule\\
  \Gamma \vdash P \Rightarrow process
  }
  {\Gamma \vdash ru; P \Rightarrow process }
\end{align*}

\begin{align*}
  \inferrule[]
  {\Gamma \vdash P_0 \Rightarrow process\\
  	\dots\\
  \Gamma \vdash P \Rightarrow process
  }
  {\Gamma \vdash choice \{ P_0, \dots \}; P \Rightarrow process }
\end{align*}

\begin{align*}
\inferrule[]
{\Gamma \vdash e \Rightarrow afact\\
	\Gamma \vdash t \Rightarrow temporal
}
{\Gamma \vdash e @ t \Rightarrow formula
}
\end{align*}

\begin{align*}
\inferrule[]
{\Gamma \vdash e_1 \Rightarrow formula\\
 \Gamma \vdash e_2 \Rightarrow formula
}
{\Gamma \vdash e_1 \& e_2 \Rightarrow formula
}
\end{align*}

\begin{align*}
	\inferrule[]
	{\Gamma \vdash e_1 \Rightarrow formula\\
		\Gamma \vdash e_2 \Rightarrow formula
	}
	{\Gamma \vdash e_1 | e_2 \Rightarrow formula
	}
\end{align*}

\begin{align*}
  \inferrule[]
  {\Gamma, x_1 \Rightarrow bits, \dots, y_1 \Rightarrow temporal, \dots
  \vdash e \Rightarrow formula
  }
  {\Gamma \vdash \text{All $x_1 \dots y_1 \dots$ \textbf{.} e} \Rightarrow formula}
\end{align*}

\begin{align*}
\inferrule[]
{\Gamma, x_1 \Rightarrow bits, \dots, y_1 \Rightarrow temporal, \dots
	\vdash e \Rightarrow formula
}
{\Gamma \vdash \text{Ex $x_1 \dots y_1 \dots$ \textbf{.} e} \Rightarrow formula}
\end{align*}

\section{Full CFG construction}
\label{appendix:cfg-construction-full}

Reader may wish to refer to
the implementation in \texttt{src/tg\_graph.ml}
when reading this section.

In the following, given a CFG $G$, we write $V(G)$ and $E(G)$ to denote its set of vertices and edges, respectively. We use the symbols $fst(X)$ and $snd(X)$ to denote the first and the second projection of $X$, respectively. 

\begin{definition} \label{def:basic-graph-functions}
  For a CFG $\langle V, E \rangle$ define:
\begin{itemize}
  \item $roots(E)$ is the
  set of labels $k_1, k_2, \dots$ such that
  $\forall i . \neg \exists k' . (k', k_i) \in E$.
  \item $succ(E, k)$ is the
  set of labels $S$ such that
  $\forall k' . (k, k') \in E \iff k' \in S$.
  \item $leaves(E)$ is the
  set of labels $k_1, k_2, \dots$ such that
  $\forall i . \neg \exists k' . (k_i, k') \in E$.
\end{itemize}
\end{definition}

\begin{definition}
  Let $P$ be a process and $k$ a label.
  The core CFG of $P$ with respect to
  $k$ is defined
  recursively by
  \begin{itemize}
    \item (\textsc{NullCfg})
    If $P = 0$ then  $cfg'(k, P) = \langle \emptyset, \emptyset \rangle.$

    \item (\textsc{ChainCfg}) 
    If $P = l \inbrac a \outbrac r; P'$ then
    \begin{align*}
      cfg'(k&, P) \\
      &=
      \big\langle
        \{ (k', l\inbrac a \outbrac r)  \}
        \cup
        V(cfg'(k', P')), \\
        &\phantom{= \big\langle }
        \{ (k, k') \}
        \cup
        E(cfg'(k', P'))
        \big\rangle
    \end{align*}
      where $k'$ is a globally fresh label
    \item (\textsc{ChoiceCfg})
    If $P = choice \{ P_1; P_2; \ldots, P_n \}; P'$ then
    \begin{align*}
      cfg'&(k, P) \\
        = \Big\langle
            &\bigcup_{i=1}^n V(cfg'(k, P_i))
            \cup
              V(cfg'(k', P')) \\
        &\hspace{8.5em}\cup
        \{ (k', []\inbrac \outbrac []) \}, \\
        &\bigcup_{i=1}^n
        \{ (k, k_i) \}
        \cup \bigcup_{i=1,j=1}^n,m \{ (l_{ij}, k') \}\\
        \cup
        &\bigcup_{i=1}^n
        E(cfg'(k, P_i))
        \cup
        E(cfg'(k', P'))
        \Big\rangle
    \end{align*}
    where $k', k_1, k_2, \ldots$ are globally fresh labels
    and $\{ l_{i1}, l_{i2}, \ldots, l_{im} \} = leaves(cfg'(k, P_i))$.
    
    Note that $(k', []\inbrac \outbrac [])$ is used to represent the case where the choices failed (in which case, $P'$ would not be executed). It can be seen as a skip command that does nothing.
  \end{itemize}
\end{definition}

To accommodate breaks and continues,
we actually need some additional
parameters for $cfg'$:
\begin{itemize}
    \item
    \textit{\textbf{Loop stack}} ($lp_{stack}$):
    Let loop skeleton $(k_T, k_F, k_A)$
    be a triple where $k_T$ refers
    to the ``true branch label,''
    $k_F$ the ``false branch label,''
    and $k_A$ the ``after loop label.''

    Loop stack is then defined as
    a list of loop skeletons.
    
    \item
    \textit{\textbf{Loop table}} ($lp_{table}$)
    is defined as a mapping
    from strings to loop skeletons
\end{itemize}
Above parameters are simply passed
as is for (\textsc{ChainCfg}),
(\textsc{ChoiceCfg})
(\textsc{If-then-elseCfg})
however,
and as such were not shown.

\begin{itemize}
    \item (\textsc{If-then-elseCfg})
    If $P = \text{$if$ $cond$ $then$} \{ P_T \} else \{ P_F \}; P'$
    and $cond = \text{$'c$ $cas$ $t$}$ then
    \begin{align*}
        cfg'&(k, P) \\
          = \Big\langle
          &\{ (k_T, [\text{$'c$ $cas$ $t$}]\inbrac\outbrac[]),
             (k_F, []\inbrac Restrict \outbrac[]),\\
             &\phantom{0} (k', []\inbrac\outbrac[]) \}\\
          &\cup V(cfg'(k_T, P_T))
          \cup V(cfg'(k_F, P_F)) \\
          &\cup \bigcup_{i=1}^n \{ (l_{Ti}, k') \}
          \cup \bigcup_{i=1}^m \{ (l_{Fi}, k') \}
          \cup \{ (k, k_T), (k, k_F) \}\\
          &\cup E(cfg'(k_T, P_T))
          \cup E(cfg'(k_F, P_F))\\
          &\cup E(cfg'(k', P'))
          \Big\rangle
    \end{align*}
    where
    $k', k_T, k_F, \ldots$ are globally fresh labels,
    $\{ l_{T1}, l_{T2}, \ldots, l_{Tn} \} = leaves(cfg'(k, P_T))$,
    $\{ l_{F1}, l_{F2}, \ldots, l_{Fm} \} = leaves(cfg'(k, P_F))$,
    and $Restrict = Neq('c, t)$ if $t$ has no variables.
    
    Otherwise $Restrict = R('c, t)$ where $R$ is a
    fresh fact name associated
    with the following restriction
    (\texttt{vars(t)} expands into variables in $t$):
    \begin{lstlisting}[style=listing0]
"All #i cell vars(t) .
  ((R(cell, t) @ #i) ==> (not (cell = t)))"
    \end{lstlisting}
    A restriction is necessary
    for this case, as
    using just $Neq$ here yields
    an ill-formed rule which
    contains free variables.
    
    Constructions for
    other cases of $cond$ are similarly
    defined.
    
    \item (\textsc{WhileCfg})
    If $P = \text{$while$ $cond$ } \{ P_T \}; P'$
    and $cond = \text{$'c$ $cas$ $t$}$ then
    \begin{align*}
        cfg'&(k, lp_{stack}, lp_{table}, P) \\
          =
          &\text{let $lp_{stack}' = (k_T, k_F, k') :: lp_{stack}$ in}\\
          \Big\langle
          &\{ (k_T, [\text{$'c$ $cas$ $t$}]\inbrac\outbrac[]),
             (k_F, []\inbrac Restrict \outbrac[]),\\
             &\phantom{0} (k', []\inbrac\outbrac[]) \}\\
          &\cup V(cfg'(k_T, lp_{stack}', lp_{table}, P_T))\\
          &\cup V(cfg'(k', lp_{stack}, lp_{table}, P')) \\
          &\cup \bigcup_{i=1}^n \{ (l_{Ti}, k_T) \}
          \cup \bigcup_{i=1}^n \{ (l_{Ti}, k_F) \}\\
          &\cup \{ (k, k_T), (k, k_F), (k_F, k') \}\\
          &\cup E(cfg'(k_T, lp_{stack}', lp_{table}, P_T))\\
          &\cup E(cfg'(k', lp_{stack}, lp_{table}, P'))
          \Big\rangle
    \end{align*}
    where
    $k', k_T, k_F, \ldots$ are globally fresh labels,
    $\{ l_{T1}, l_{T2}, \ldots, l_{Tn} \} = leaves(cfg'(k, lp_{stack}', lp_{table}, P_T))$,
    and $Restrict$ is defined in the same way
    as in (\textsc{If-then-elseCfg}).
    
    Constructions for
    other cases of $cond$ are similarly
    defined.
    
    \item (\textsc{Labeled-WhileCfg})
    If $P = \text{``$lp_{label}$'': $while$ $cond$ } \{ P_T \}; P'$
    and $cond = \text{$'c$ $cas$ $t$}$ then
    \begin{align*}
        cfg'&(k, lp_{stack}, lp_{table}, P) \\
          =
          &\text{let $lp_{table}' = \{ \langle lp_{label}, (k_T, k_F, k') \rangle \} \cup lp_{table}$ in}\\
          &\dots
    \end{align*}
    the $\dots$ and the remainder are the same
    as the definitions in (\textsc{WhileCfg})
    except $lp_{table}'$ is used when
    recursing into $P_T$.

    \item (\textsc{LoopCfg})
    If $P = \text{$loop$ } \{ P_{lp} \}; P'$ then
    \begin{align*}
        cfg'&(k, lp_{stack}, lp_{table}, P) \\
          =
          &\text{let $lp_{stack}' = (k_{lp}, NULL, k') :: lp_{stack}$ in}\\
          \Big\langle
          &\{ (k', []\inbrac\outbrac[]) \}\\
          &\cup V(cfg'(k_{lp}, lp_{stack}', lp_{table}, P_{lp}))\\
          &\cup V(cfg'(k', lp_{stack}, lp_{table}, P')) \\
          &\cup \bigcup_{i=1}^n \{ (l_{Ti}, k_{lp}) \}
          \cup \bigcup_{i=1}^n \{ (l_{Ti}, k_F) \}\\
          &\cup \{ (k, k_{lp}) \}\\
          &\cup E(cfg'(k_{lp}, lp_{stack}', lp_{table}, P_{lp}))\\
          &\cup E(cfg'(k', lp_{stack}, lp_{table}, P'))
          \Big\rangle
    \end{align*}
    where
    $k', k_{lp}$ are globally fresh labels,
    $\{ l_{T1}, l_{T2}, \ldots, l_{Tn} \} = leaves(cfg'(k, lp_{stack}', lp_{table}, P_T))$,
    and $Restrict$ is defined in the same way
    as in (\textsc{If-then-elseCfg}).
    
    \item (\textsc{Labeled-LoopCfg})
    If $P = \text{``$lp_{label}$'': $loop$ } \{ P_{lp} \}; P'$
    and $cond = \text{$'c$ $cas$ $t$}$ then
    \begin{align*}
        cfg'&(k, lp_{stack}, lp_{table}, P) \\
          =
          &\text{let $lp_{table}' = \{ \langle lp_{label}, (k_{lp}, NULL, k') \rangle \} \cup lp_{table}$ in}\\
          &\dots
    \end{align*}
    the $\dots$ and the remainder are the same
    as the definitions in (\textsc{LoopCfg})
    except $lp_{table}'$ is used when
    recursing into $P_{lp}$.

    \item (\textsc{BreakCfg})
    If $P = \text{$break$}$ then
    \begin{align*}
        cfg'&(k, lp_{stack}, lp_{table}, P) \\
          =
          &\text{let $(k_T, k_F, k_A), \dots = lp_{stack}$ in}\\
          \Big\langle
          & \emptyset, \{ (k, k_A) \}
          \Big\rangle
    \end{align*}

    \item (\textsc{Labeled-BreakCfg})
    If $P = \text{$break$ ``$lp_{label}$''}$ then
    \begin{align*}
        cfg'&(k, lp_{stack}, lp_{table}, P) \\
          =
          &\text{let $(k_T, k_F, k_A) = find(lp_{label}, lp_{table})$ in}\\
          \Big\langle
          & \emptyset, \{ (k, k_A) \}
          \Big\rangle
    \end{align*}
    
    \item (\textsc{ContinueCfg})
    If $P = \text{$continue$}$ then
    \begin{align*}
        cfg'&(k, lp_{stack}, lp_{table}, P) \\
          =
          &\text{let $(k_T, k_F, k_A), \dots = lp_{stack}$ in}\\
          \Big\langle
          & \emptyset, \text{$if$ $k_F = NULL$ $then$ } \{ (k, k_T) \}\\
          &\text{ $else$ } \{ (k, k_T), (k, k_F) \}
          \Big\rangle
    \end{align*}
    
    \item (\textsc{Labeled-ContinueCfg})
    If $P = \text{$continue$ ``$lp_{label}$''}$ then
    \begin{align*}
        cfg'&(k, lp_{stack}, lp_{table}, P) \\
          =
          &\text{let $(k_T, k_F, k_A) = find(lp_{label}, lp_{table})$ in}\\
          \Big\langle
          & \emptyset, \text{$if$ $k_F = NULL$ $then$ } \{ (k, k_T) \} \\
          &\text{ $else$ } \{ (k, k_T), (k, k_F) \}
          \Big\rangle
    \end{align*}
\end{itemize}
 
Next, we define the CFG of a process and a system.

\begin{definition} \label{def-cfg-fun}
  The CFG of a process $P$ is given by
  prefixing the core CFG with
  a rule to generate fresh process ID
  \begin{align*}
    cfg(P)
       = \big\langle
         &\{ (k, [Fr(\sim pid)]\inbrac \outbrac ['pid := \sim pid ]) \} \\
         &\cup V(cfg'(k', P)), \\
      &\{ (k, k') \} \cup E(cfg'(k', P))
         \big\rangle
  \end{align*}
  where $k, k'$ are a globally fresh label.
\end{definition}

\begin{definition}
The CFG representation of a system,
which is a set of processes $P_1, P_2, \dots, P_n$,
is defined as:
\begin{align*}
  \Big\langle
  \bigcup_{i=1}^n V(cfg(P_i)),
  \bigcup_{i=1}^n E(cfg(P_i))
  \Big\rangle
\end{align*}
\end{definition}

\onecolumn

\section{Cell lifetime analysis is sound}

For any given trace,
if a cell is not in the final state,
there must be a rule
earlier that undefines the cell,
or the cell is not defined in the
initial state.

\begin{lemma} \label{lemma0}
For any trace $\tau = S_0, tr(S_0, [k_0, \dots, k_n]) = S_0, k_0, S_1, \dots, S_n, k_n, S_{n+1}$,
any cell $a$,
\textbf{if} $a \not \in S_{n+1}$,
\textbf{then}
\begin{align*}
(\exists i.
  (0 \leq i \leq n)
  \land
  a \in cu_u(k_i)
  \land
  (\forall j.
   (i < j \leq n)
   \Rightarrow
   a \not \in cu_d(k_j)
  )
)
\lor
a \not \in S_0
\end{align*}
\end{lemma}

{\setlength{\parindent}{0pt}
  \textbf{Proof}
  
  We prove by induction over trace $\tau$.
  
  \textbf{Base case}, $\tau = S_0,tr(S_0, []) = S_0$: the property holds trivially.
  
  \textbf{Step case}, $\tau = S_0,tr(S_0, [k_0, \dots, k_n]) = S_0, k_0, S_1, \dots, S_n, k_n, S_{n+1}$:
  \begin{adjustwidth}{1em}{}
    We assume arbitrary cell $a$ such that $a \not \in S_{n+1}$.
    By definition of well-formed trace, $S_{n+1} = (S_n \cup cu_d(k_n)) \backslash cu_u(k_n)$,
    where sets $cu_d(k_n)$ and $cu_u(k_n)$ are disjoint.

    \begin{adjustwidth}{1em}{}
      \textbf{Case} $a \in cu_d(k_n)$ and $a \in cu_u(k_n)$: contradicts the premise that the two sets are disjoint.
      
      \textbf{Case} $a \in cu_d(k_n)$ and $a\not\in cu_u(k_n)$: we have $a\in S_{n+1}$, contradicting our
      assumption.
      
      \textbf{Case} $a \not \in cu_d(k_n)$ and $a \in cu_u(k_n)$: the property holds as $a\in cu_u(k_i)$
      for some $0 \leq i \leq n+1$,
      and there does not exist $j$ such that $i < j < n$.
      
      \textbf{Case} $a \not \in cu_d(k_n)$ and $a \not \in cu_u(k_n)$:
      \begin{adjustwidth}{1em}{}
        By definition of well-formed trace, $a \not \in S_n$.
        By induction hypothesis,
        either $\exists i'. (0 \leq i' \leq n-1)
        \land
        a \in cu_u(k_{i'})
        \land
        (\forall j' . (i' < j' \leq n-1) \Rightarrow a \not\in cu_d(k_{j'}))$, or $a \not\in S_0$.

        \begin{adjustwidth}{1em}{}
          \textbf{Case} $\exists i'. (0 \leq i' \leq n-1)
        \land
        a \in cu_u(k_{i'})
        \land
        (\forall j' . (i' < j' \leq n-1) \Rightarrow a \not\in cu_d(k_{j'}))$:
          since $a\not\in cu_d(k_n)$,
          we have $\forall j' . (i' < j' \leq n) \Rightarrow a \not \in cu_d(k_{j'})$. And since $0 \leq i' \leq n-1$ implies $0 \leq i' \leq n$, the property holds for $\tau$.
          
          \textbf{Case} $a \not\in S_0$: property holds for $\tau$ trivially.
        \end{adjustwidth}
      \end{adjustwidth} 
    \end{adjustwidth}
    
  \end{adjustwidth}
$\square$
}

If a cell $a$ was undefined at any point,
and no later rules add $a$ back into the state,
then $a$ is not defined at the final state.

\begin{lemma} \label{lemma1}
  For any trace $\tau = S_0, tr(S_0, [k_0, \dots, k_n]) = S_0, k_0, S_1, \dots, S_n, k_n, S_{n+1}$,
  any cell $a$, \textbf{if}
  \begin{align*}
    \exists i.
    (0 \leq i \leq n)
    \land
    a \in cu_u(k_i)
    \land
    (\forall j.
    (i < j \leq n)
    \Rightarrow
    a \not \in cu_d(k_j)
    )
  \end{align*}
  \textbf{then} $a \not \in S_{n+1}$
\end{lemma}

{\setlength{\parindent}{0pt}
  \textbf{Proof}
  
  We prove by induction over trace $\tau$.
  
  \textbf{Base case}, $\tau = S_0, tr(S_0, []) = S_0$: the property holds vacuously.
  
  \textbf{Step case}, $\tau = S_0, tr(S_0, [k_0, \dots, k_n]) = S_0, k_0, S_1, \dots, S_n, k_n, S_{n+1}$:
  \begin{adjustwidth}{1em}{}
    We assume arbitrary cell $a$,
    and some $i$ such that
    $a \in cu_u(k_i)$
    and
    $\forall j. (i < j \leq n) \Rightarrow a\not \in cu_d(k_j)$ (1).
    \begin{adjustwidth}{1em}{}
      \textbf{Case} $0 \leq i < n$: by induction
      hypothesis, $a \not\in S_n$. By (1),
      $a \not \in cu_d(k_n)$.
      By definition of well-formed trace,
      $S_{n+1} = (S_n \cup cu_d(k_n)) \backslash cu_u(k_n)$,
      and so $a \not \in S_{n+1}$ as required.

      \textbf{Case} $i = n$:
        $a \in cu_u(k_i) = cu_u(k_n)$.
        By definition of well-formed trace,
        $S_{n+1} = (S_n \cup cu_d(k_n)) \backslash cu_u(k_n)$,
        and so $a \not \in S_{n+1}$ as required.
    \end{adjustwidth}
  \end{adjustwidth}

  $\square$
}

If a cell $a$ was not defined at $S_0$,
and no later rules added $a$ back into the state,
then $a$ is not defined at the final state.

\begin{lemma} \label{lemma2}
  For any trace $\tau = S_0, tr(S_0, [k_0, \dots, k_n]) = S_0, k_0, S_1, \dots, S_n, k_n, S_{n+1}$,
  any cell $a$, \textbf{if}
  \begin{align*}
    a \not\in S_0
    \land
    (\forall i.
    (0 \leq i \leq n)
    \Rightarrow
    a \not \in cu_d(k_i)
    )
  \end{align*}
  \textbf{then} $a \not \in S_{n+1}$.
\end{lemma}

{\setlength{\parindent}{0pt}
  \textbf{Proof}
  
  We prove by induction over
  length of trace $\tau$.
  
  \textbf{Base case}, $\tau = S_0, tr(S_0, []) = S_0$: the property holds trivially.
  
  \textbf{Step case}, $\tau = S_0, tr(S_0, [k_0, \dots, k_n]) = S_0, k_0, S_1, \dots, S_n, k_n, S_{n+1}$:
  \begin{adjustwidth}{1em}{}
    We assume arbitrary cell $a$,
    where $a \not\in S_0$,
    and $\forall i. (0 \leq i \leq n) \Rightarrow a\not\in cu_d(k_i)$ (1).
    By induction hypothesis, $a \not\in S_n$.
    By (1), $a \not \in cu_d(k_n)$. By definition
    of well-formed trace, $S_{n+1} = (S_n \cup cu_d(k_n)) \backslash cu_u(k_n)$, and so $a \not \in S_{n+1}$ as required.
  \end{adjustwidth}
  $\square$
}

Given a valid trace,
for any rule that references a cell $a$,
there must be an earlier rule defining cell $a$,
and no rules after said defining rule undefining it.

\begin{lemma} \label{lemma3}
  For any trace $\tau = S_0, tr(S_0, [k_0, \dots, k_n]) = S_0, k_0, S_1, \dots, S_n, k_n, S_{n+1}$, any cell $a$,
  \textbf{if} $\tau$ is valid $\land$ has at least one element $\land$ $a \not \in S_0$,
  \textbf{then}
  $a \not\in cu_r(k_0)$ and
  \begin{align*}
    \forall i.
    ( (0 < i \leq n) \land a \in cu_r(k_i) )
    \Rightarrow
    (\exists j.
     (0 \leq j < i)
     \land
     a \in cu_d(k_j)
     \land
     \forall l. (j < l < i)
     \Rightarrow
     a \not \in cu_u(k_l)
    )
  \end{align*}
\end{lemma}

{\setlength{\parindent}{0pt}
  \textbf{Proof}
  
  We prove by induction over trace $\tau$.
  
  \textbf{Base case}, $\tau = S_0, tr(S_0, []) = S_0$. The property holds vacuously.
  
  \textbf{Step case}, $\tau = S_0, tr(S_0, [k_0, \dots, k_n]) = S_0, k_0, S_1, \dots, S_n, k_n, S_{n+1}$:
  \begin{adjustwidth}{1em}{}
    We pick arbitrary cell $a$.
    We assume $\tau$ is valid and $a \not\in S_0$.
    
    To prove $a\not \in cu_r(k_0)$:
    \begin{adjustwidth}{1em}{}
      Suppose $a\in cu_r(k_0)$,
      since $a \not\in S_0$ by assumption,
      $\tau$ is invalid by definition, contradicting our assumption.
    \end{adjustwidth}
  
    To prove the other statement:
    \begin{adjustwidth}{1em}{}
      We pick arbitrary $i$ such that
      $0 < i \leq n$
      and
      $a\in cu_r(k_i)$.
      We want to show
      $
      \exists j.
      (0 \leq j < i)
      \land
      a \in cu_d(k_j)
      \land
      \forall l. (j < l < i)
      \Rightarrow
      a \not \in cu_u(k_l)
      $ by contradiction.
      
      We suppose the negation:
      $
      \forall j.
      (i \leq j)
      \lor
      a \not \in cu_d(k_j)
      \lor
      \exists l. (j < l < i) \land a \in cu_u(k_l)
      $ (1).
      
      We show that this implies
      $\forall j. (0 \leq j < i) \Rightarrow a \not \in cu_d(k_j) \land a \not\in cu_u(k_j)$.
      
      \begin{adjustwidth}{1em}{}
        We prove by induction over $j$ in reverse.
        
        \textbf{Base case}, $j = i - 1$:
        since there does not exist $l$ such that $j < l < i$,
        by (1), $a \not\in cu_d(k_j)$.
        Suppose $a \in cu_u(k_j)$, then we have
        $a \not \in S_i = (S_j \cup cu_d(k_j)) \backslash cu_u(k_j)$,
        and since $a \in cu_r(k_i)$ by assumption,
        this renders $\tau$ invalid, contradicting our assumption.
        So $a \not \in cu_u(k_j)$
        
        \textbf{Step case}, $0 \leq j < i - 1$:
        \begin{adjustwidth}{1em}{}
          Induction hypothesis:
          $\forall z. (j \leq z < i) \Rightarrow a \not \in cu_d(k_z) \land a \not \in cu_u(k_z)$.
          
          By (1), either $a \not \in cu_d(k_j)$
          or
          $\exists l. (j < l < i) \land a \in cu_u(k_l)$.
          
          \begin{adjustwidth}{1em}{}
            \textbf{Case} $\exists l. (j < l < i) \land a \in cu_u(k_l)$ holds:
            then we have $l$ such that $a \in cu_u(k_l)$, and $a \not \in cu_u(k_l)$ by induction hypothesis. Contradiction.
            
            \textbf{Case} $\exists l. (j < l < i) \land a \in cu_u(k_l)$ does not hold:
            then we have $a \not \in cu_d(k_j)$.
            Suppose $a \in cu_u(k_j)$,
            and since induction hypothesis implies
            $\forall z. (j \leq z \leq i - 1) \Rightarrow a \not\in cu_d(k_z)$,
            by Lemma~\ref{lemma1}, $a \not\in S_i$.
            Since $a \in cu_r(k_i)$, this means $\tau$ is invalid,
            contradicting our assumption. So $a \not\in cu_u(k_j)$.
          \end{adjustwidth}
        \end{adjustwidth}
      \end{adjustwidth}
      
      From $\forall j. (0 \leq j < i) \Rightarrow a \not \in cu_d(k_j) \land a \not\in cu_u(k_j)$
      we have $\forall j. (0 \leq j \leq i - 1) \Rightarrow a \not \in cu_d(k_j)$
      and by assumption we have $a \not \in S_0$.
      So by Lemma~\ref{lemma2}, we have $a \not \in S_i$.
      
      But since $a \in cu_r(k_i)$ by assumption, this means $\tau$ is invalid, contradicting our assumption.
      
    \end{adjustwidth}
  \end{adjustwidth}
  $\square$
}

\begin{lemma} \label{lemma4}
  For any trace $\tau = S_0, tr(S_0, [k_0, \dots, k_n]) = S_0, k_0, S_1, \dots, S_n, k_n, S_{n+1}$,
  any cell $a$,
  \textbf{if}
  \begin{align*}
    \exists i. (0 \leq i \leq n)
    \land
    a \in cu_d(k_i)
    \land
    \forall j. (i < j \leq n) \Rightarrow a\not\in cu_u(k_j)
  \end{align*}
  \textbf{then} $a\in S_{n+1}$
\end{lemma}

{\setlength{\parindent}{0pt}
  \textbf{Proof}
  
  We prove by induction over $\tau$.
  
  \textbf{Base case}, $\tau = S_0, tr(S_0, []) = S_0$:
  the property holds vacuously.
  
  \textbf{Step case}, $\tau = S_0, tr(S_0, [k_0, \dots, k_n]) = S_0, k_0, \dots, S_n, k_n, S_{n+1}$
  \begin{adjustwidth}{1em}{}
    Suppose there exists $i$
    where
    $0 \leq i \leq n$
    and
    (1) $a \in cu_d(k_i)$
    and
    (2) $\forall j. (i < j \leq n) \Rightarrow a \not \in cu_u(k_j)$.
    
    \textbf{Case} $0 \leq i < n$:
    \begin{adjustwidth}{1em}{}
      Then we have $i$ where $0 \leq i \leq n-1$.
      And since
      (2)
      implies
      $\forall j. (i < j \leq n-1) \Rightarrow a \not\in cu_u(k_j)$,
      by induction hypothesis $a \in S_n$.
      By (2),
      we also have $a \not\in cu_u(k_n)$.
      By definition of
      well-formed trace, $S_{n+1} = (S_n \cup cu_d(k_n)) \backslash cu_u(k_n)$,
      and so $a \in S_{n+1}$.
    \end{adjustwidth}
    
    \textbf{Case} $i = n$:
    by (1) we have $a \in cu_d(k_n)$
    and (2) we have $a \not \in cu_u(k_n)$.
    By definition of well-formed trace, $S_{n+1} = (S_n \cup cu_d(k_n)) \backslash cu_u(k_n)$,
    and so $a \in S_{n+1}$.
  \end{adjustwidth}
  $\square$
}

\begin{definition}    
We make use of a shorthand for retrieving the last state of trace:
    \begin{align*}
      S_{end}([S_0, k_0, \dots, k_n, S_{n+1}]) = S_{n+1}
    \end{align*}
\end{definition}

\begin{definition}
  We define function $find_{tr}$ which retrieves
  the index of an appearance of rule indexed by $k$
  in a trace as follows:
  \begin{align*}
    find_{tr}(k, \tau) = i
  \end{align*}
 such that $k_i = k$, and $k_i$ appears at the $i$th position in trace $\tau$.
 
 This helps us talk about ``copies'' of the same rule across
 different traces in the remainder of our proofs.
\end{definition}

\begin{lemma} \label{lemma5}
\setlength{\parindent}{0pt}

For all ids
$P_{ids} = [p_0, \dots, p_a]$,
$X_{ids} = [x_0, \dots, x_b]$,
$Y_{ids} = [y_0, \dots, y_c]$,
$Z_{ids} = [z_0, \dots, z_d]$,
for all $S_0$,
\textbf{if} the following traces are valid:
\begin{itemize}
  \item $S_0, P, X, Y_1$
  \item $S_0, P, X, Z_1$
  \item $S_0, P, Y_2, Z_2$
\end{itemize}
where
\begin{align*}
  P &= tr(S_0, P_{ids})\\
  X &= tr(S_{end}(P), X_{ids})\\
  Y_1 &= tr(S_{end}(X), Y_{ids})\\
  Z_1 &= tr(S_{end}(X), Z_{ids})\\
  Y_2 &= tr(S_{end}(P), Y_{ids})\\
  Z_2 &= tr(S_{end}(Y_2), Y_{ids})
\end{align*}
\textbf{then} trace $S_0, P, X, Y_1, Z_3$ is also valid
where $Z_3 = tr(S_{end}(Y_1), Z_{ids})$

\end{lemma}

{\setlength{\parindent}{0pt}
  \textbf{Proof}

  We pick arbitrary ids
  $P_{ids} = [p_0, \dots, p_a]$,
  $X_{ids} = [x_0, \dots, x_b]$,
  $Y_{ids} = [y_0, \dots, y_c]$,
  $Z_{ids} = [z_0, \dots, z_d]$,
  and arbitrary initial state $S_0$.
  
  We assume traces
  $[S_0, P, X, Y_1]$,
  $[S_0, P, X, Z_1]$,
  $[S_0, P, Y_2, Z_2]$
  are valid
  (copying the definitions of the lemma).
  
  Now we show trace $S_0, P, X, Y_1, Z_3$ (same
  definition as lemma) is also valid by contradiction.
  
  Suppose trace $S_0, P, X, Y_1, Z_3$ is not valid,
  then by definition there exists rule
  indexed by $k_i$, and cell $a \in cu_r(k_i)$
  where $a \not \in S_i$.
  Furthermore, we know $k_i \in Z_4$
  and $k_i \in Z_{ids}$,
  since if $k_i$ is in
  $P, X, Y_1$,
  then trace $S_0, P, X, Y_1$ would be invalid,
  contradicting our assumption.
  
  Since $a \not \in S_i$,
  by Lemma~\ref{lemma0},
  either (1) there exists $j$,
  where $0 \leq j \leq i-1$
  and $a \in cu_u(k_j)$
  and $\forall l. (j < l \leq i-1) \Rightarrow a \not\in cu_d(k_l)$,
  or (2) $a \not \in S_0$.

  \begin{adjustwidth}{1em}{}
    \textbf{Case} (1) holds:
    \begin{adjustwidth}{1em}{}
      \textbf{Case} $k_j \in Z_3$:
      \begin{adjustwidth}{1em}{}
        Then we have $k_j \in Z_{ids}$,
        and as such $k_j \in Z_3$ as well.
        Let $i' = find_{tr}(k_i, [S_0, P, X, Z_1])$
        and $j' = find_{tr}(k_j, [S_0, P, X, Z_1])$.
        
        Since $a \not \in cu_d(k_l)$ for all
        rules $k_l$ between $k_j$ and $k_i$,
        we also have $a \not \in cu_d(k_l')$
        for all rule $k_l'$ between $k_j'$ and $k_i'$.
        By Lemma~\ref{lemma1} trace $S_0, P, X, Z_1$
        (or any trace with $Z_i$ for some $i$ as suffix)
        would also be invalid,
        which
        contradicts our assumption.
      \end{adjustwidth}
      
      \textbf{Case} $k_j \in Y_1$:
      \begin{adjustwidth}{1em}{}
        Then we have $k_j \in Y_{ids}$,
        and as such $k_j = k_j \in Y_2$ as well.
        Let $i' = find_{tr}(k_i, [S_0, P, Y_2, Z_2])$
        and $j' = find_{tr}(k_j, [S_0, P, Y_2, Z_2])$.
        
        Since $a \not \in cu_d(k_l)$ for all
        rules $k_l$ between $k_j$ and $k_i$,
        we also have $a \not \in cu_d(k_l')$
        for all rule $k_l'$ between $k_j'$ and $k_i'$.
        By Lemma~\ref{lemma1} trace $S_0, P, Y_2, Z_2$
        would also be invalid,
        which
        contradicts our assumption.
      \end{adjustwidth}
      
      \textbf{Case} $k_j \in X$:
      \begin{adjustwidth}{1em}{}
        Let $i' = find_{tr}(k_i, [S_0, P, X, Z_1])$.
        
        Since $a \not \in cu_d(k_l)$ for all
        rules $k_l$ between $k_j$ and $k_i$,
        we also have $a \not \in cu_d(k_{l'})$
        for all rule $k_{l'}$ between $k_j$ and $k_{i'}$.
        By Lemma~\ref{lemma1} trace $S_0, P, X, Z_1$
        would also be invalid,
        which
        contradicts our assumption.
      \end{adjustwidth}

      \textbf{Case} $k_j \in P$:
      \begin{adjustwidth}{1em}{}
        Let $i' = find_{tr}(k_i, [S_0, P, X, Z_1])$.
        
        Since $a \not \in cu_d(k_l)$ for all
        rules $k_l$ between $k_j$ and $k_i$,
        we also have $a \not \in cu_d(k_l')$
        for all rules $k_l'$ between $k_j$ and $k_i'$.
        By Lemma~\ref{lemma1} trace $S_0, P, X, Z_1$
        would also be invalid,
        which
        contradicts our assumption.
      \end{adjustwidth}
    \end{adjustwidth}
  
    \textbf{Case} (2) holds:
    \begin{adjustwidth}{1em}{}
      
      Since $S_0, P, X, Z_1$ is valid:
      \begin{adjustwidth}{1em}{}
        Since $k_i \in Z_{ids}$, we have $k_i \in Z_1$.
        Let $i' = find_{tr}(k_i, [S_0, P, X, Z_1])$.
        
        By Lemma~\ref{lemma3}, we have $j'$
        where $0 \leq j' < i'$
        and
        $a \in cu_d(k_{j'})$
        and
        $\forall l'. (j' < l' < i') \Rightarrow a\notin cu_u(k_{l'})$.
        
        Then we have the following possible cases:
        
        \begin{adjustwidth}{1em}{}
          \textbf{Case} (A1) $k_{j'} \in Z_1$:
          \begin{adjustwidth}{1em}{}
            Then there exists
            a subsequence in $Z_1$
            such that $a$ is defined at the
            beginning of the subsequence,
            and
            no rules undefine $a$ up to
            rule $k_{i'}$ which references $a$.
          \end{adjustwidth}
        
          \textbf{Case} (A2) $k_{j'} \in X$:
          \begin{adjustwidth}{1em}{}
            Then there exists
            a suffix of $X$
            such that $a$ is defined at the beginning
            of the suffix and
            no rules undefine $a$ up to
            rule $k_{i'}$ which references $a$.
            
            And there exists
            a prefix of $Z_1$ such
            that
            no rules undefine $a$ up to rule $k_{i'}$.
          \end{adjustwidth}
        
          \textbf{Case} (A3) $k_{j'} \in P$:
          \begin{adjustwidth}{1em}{}
            Then there exists a suffix of $P$
            such that $a$ is defined
            at the beginning of the suffix and
            remains well-defined for the entire suffix.
            
            And no rules undefine $a$ in $X$.
            
            And there exists a prefix
            of $Z_1$ such that
            no rules undefine $a$ up to rule $k_{i'}$.
          \end{adjustwidth}
        \end{adjustwidth}
      \end{adjustwidth}
    
      Since $S_0, P, Y_2, Z_2$ is valid:
      \begin{adjustwidth}{1em}{}
        Since $k_i \in Z_{ids}$, we have $k_i \in Z_2$.
        Let $i' = find_{tr}(k_i, [S0, P, Y_2, Z_2])$.
        
        By Lemma~\ref{lemma3}, we have $j'$
        where $0 \leq j' < i'$
        and
        $a \in cu_d(k_{j'})$
        and
        $\forall l'. (j' < l' < i') \Rightarrow a\notin cu_u(k_{l'})$.
        
        Then we have the following possible cases:
        \begin{adjustwidth}{1em}{}
          \textbf{Case} (B1) $k_{j'} \in Z_2$:
          \begin{adjustwidth}{1em}{}
            Then there exists
            a subsequence in $Z_2$
            such that $a$ is defined at the
            beginning of the subsequence,
            and
            no rules undefine $a$ up to
            rule $k_{i'}$.
          \end{adjustwidth}

          \textbf{Case} (B2) $k_{j'} \in Y_2$:
          \begin{adjustwidth}{1em}{}
            Then there exists
            a suffix of $Y_2$
            such that $a$ is defined at the beginning
            of the suffix and
            no rules undefine $a$ up to
            rule $k_{i'}$ which references $a$.
            
            And there exists
            a prefix of $Z_2$ such
            that
            no rules undefine $a$ up to rule $k_{i'}$.
          \end{adjustwidth}
        
          \textbf{Case} (B3) $k_{j'} \in P$:
          \begin{adjustwidth}{1em}{}
            Then there exists a suffix of $P$
            such that $a$ is defined
            at the beginning of the suffix and
            remains well-defined for the entire suffix.
            
            And no rules undefine $a$ in $Y_2$.
            
            And there exists a prefix
            of $Z_2$ such that
            no rules undefine $a$ up to rule $k_{i'}$.
          \end{adjustwidth}
        \end{adjustwidth}
      \end{adjustwidth}
    
      Overall we have $(\text{A1} \lor \text{A2} \lor \text{A3}) \land (\text{B1} \lor \text{B2} \lor \text{B3})$.
      We case analyze as follows:
      \begin{adjustwidth}{1em}{}
        \textbf{Case} (A1) or (B1):
        \begin{adjustwidth}{1em}{}
          Then we have $k_{j'} \in Z_{ids}$ for some $k_{j'}$,
          and in turn $k_{j'} \in Z_4$.
          Let $j = find_{tr}(k_{j'}, [S_0, X, Y_1, Z_4])$.
          
          Since no rules between $k_{j'}$ and $k_{i'}$ undefine
          $a$, similarly no rules between $k_j$ and $k_i$
          undefine $a$.
          
          By Lemma~\ref{lemma4}, $a \in S_i$, contradicting
          assumption of trace $S_0, P, X, Y_1, Z_4$ being invalid.
        \end{adjustwidth}
      
        \textbf{Case} (B2):
        \begin{adjustwidth}{1em}{}
          Then we have $k_{j'} \in Y_{ids}$ for some $k_{j'}$,
          and in turn $k_{j'} \in Y_1$.
          Let $j = find_{tr}(k_{j'}, [S_0, X, Y_1, Z_4])$.
          
          Since no rules between $k_{j'}$ and $k_{i'}$ undefine
          $a$, similarly no rules between $k_j$ and $k_i$
          undefine $a$.
          
          By Lemma~\ref{lemma4}, $a \in S_i$, contradicting
          assumption of trace $S_0, P, X, Y_1, Z_4$ being invalid.
        \end{adjustwidth}
      
        \textbf{Case} (A2) and (B3):
        \begin{adjustwidth}{1em}{}
          Then we have $k_j \in X$ for some $k_j$.
          And prefix of $Z_4$ such that no
          rules undefine $a$ up to $k_i$.
          
          From (B3), no rules in $Y_{ids}$ undefine $a$,
          and in turn no rules in $Y_1$ undefine $a$.
          
          Overall this means no rules between $k_j$
          and $k_i$ undefine $a$.
          
          By Lemma~\ref{lemma4}, $a \in S_i$,
          contradicting assumption of trace
          $S_0, P, X, Y_1, Z_4$ being invalid.
        \end{adjustwidth}
      
        \textbf{Case} (A3) and (B3):
        \begin{adjustwidth}{1em}{}
          Then we have $k_j \in P$ for some $k_j$.
          And prefix of $Z_4$ such that no
          rules undefine $a$ up to $k_i$.
          
          From (A3), no rules in $X$ undefine $a$.
          
          From (B3), no rules in $Y_{ids}$ undefine $a$,
          and in turn no rules in $Y_1$ undefine $a$.
          
          Overall this means no rules between $k_j$
          and $k_i$ undefine $a$.
          
          By Lemma~\ref{lemma4}, $a \in S_i$,
          contradicting assumption of trace
          $S_0, P, X, Y_1, Z_4$ being invalid.
        \end{adjustwidth}
      
      \end{adjustwidth}
    
    \end{adjustwidth}
  \end{adjustwidth}
  $\square$
}

\section{Maximal context inference is correct}

\begin{definition}
  \setlength{\parindent}{0pt}
  
  An execution trace of a CFG $\langle V, E \rangle$,
  which we call CFG trace below to avoid
  confusion with (actual) execution trace of a Tamgram process,
  is an alternating sequence between
  a set of cells and
  and rule ids, e.g.~$S_0, k_0, S_1, k_1, \dots$,
  where $S_n$ denotes the set of cells which are defined
  up to position $n$ in the trace,
  $(k_n, k_{n+1}) \in E$.
\end{definition}

\begin{definition}
The cell usage of a MSR rule
$ru = l\inbrac a\outbrac r$
is defined as a triple $cu(ru) = \langle r, d, u \rangle$, where
\begin{itemize}
  \item $r =$ references of cells in any of $l, a, r$
  \item $d =$ definitions of cells by \verb|:=| assignment syntax in $r$
  \item $u =$ undefining of cells by reserved function symbol \verb|undef|
\end{itemize}

Given a CFG $\langle V, E \rangle$,
we define the following short-hands for
a MSR rule $ru_k$ labeled by $k$,
i.e. $(k, ru_k) \in V$.
Let $\langle r, d, u \rangle = cu(ru_k)$, then we define: 
\[ 
\begin{array}{ll}
  cu(k) = cu(ru_k) & cu_r(ru_k) = r\\
  cu_d(ru_k) = d & cu_u(ru_k) = u
\end{array}
\]

We say a rule $ru$ references (or requires) cell $c$ if and only if
$c \in cu_r(ru)$ etc.
\end{definition}

We say a CFG trace is well-formed when the states are ``coherent''
and no rules define and undefine the same cells
simultaneously.

\begin{definition}
We define the function which computes the next
state in a CFG trace as:
\begin{align*}
    next(S, k) = (S \cup cu_d(k)) \backslash cu_u(k)
\end{align*}
where $S$ is the current state, and $k$ is the
index of the rule chosen to make progress.
\end{definition}

\begin{definition}
  A CFG trace is well-formed if and only if:
    \begin{align*}
      \forall n. S_{n+1} = next(S_n, k_n)
    \end{align*}
  and
    \begin{align*}
        \forall n. cu_d(k_n) \cap cu_u(k_n) = \emptyset
    \end{align*}
\end{definition}

\begin{example}
Suppose rule $0$ defines cell $a$: $[In(x)]\inbrac~\outbrac ['a := x]$, rule $1$
defines cell $b$ but undefines $a$: $[In(y)]\inbrac~\outbrac ['b := y, undef('a)]$,
then an example of a well-formed execution trace would be as follows:
\begin{align*}
  \emptyset, 0, \{ a \}, 1, \{ b \}
\end{align*}
\end{example}

We also define a shorthand for constructing a well-formed CFG trace
from an initial state $S_0$ and a sequence of rules:

\begin{definition}
    Given a set of cells $S_0$ and a list
    of step labels, $tr$ constructs a partial trace recursively:
    \begin{align*}
      tr(S_0, []) &= []\\
      tr(S_0, [k_0, \dots, k_n]) &= k_0, S_1, k_1, \dots, S_n, k_n, S_{n+1}
    \end{align*}
  where $S_{n+1} = next(S_n, k_n)$.
\end{definition}
$S_0$ is not part of $tr(S_0, \dots)$
as this simplifies certain syntactic
definitions in proofs.

\begin{remark}
  Strictly speaking, a CFG trace 
  consists of instances of rules. However, since instantiations
  of rules do not affect the discussion of cell usage,
  i.e.~$cu(ru) = cu(\sigma(ru))$ for any variable substitution $\sigma$,
  we use the above simpler definitions.
\end{remark}

\begin{definition}
  A CFG trace is invalid if
  $cu_r(k_n) \not \subseteq S_n$ for some $n$, i.e.~the rule references
  an undefined cell.
  Conversely a trace is valid when
  $cu_r(k_n) \subseteq S_n$ for all $n$.
\end{definition}

\begin{remark}
  In principle a trace can be not well-formed but valid,
  but we are only interested in traces which are well-formed.
  As such we use ``CFG trace'' and ``well-formed CFG trace'' interchangeably
  in the remainder of this paper.
\end{remark}

\begin{definition}
  We say a CFG trace $S_0, k_0, \dots$ is loop-free
  iff
  for any index $k$, $k$ appears at most once in the trace.
  More generally, we say a trace has maximum of $n$ loops (or up to $n$ loops) iff
  for any index $k$, $k$ appears at most $n+1$ times in the trace.
  This matches the usual definition of loop-free path.
\end{definition}

\begin{example}
An illustration of a trace that is not loop-free,
where rule 2 appears twice in the trace:
\begin{align*}
\emptyset,
0,
\{ a \},
1,
\{ b \},
2,
\{ b, c \},
3,
\{ b, c, d \},
4,
\{ b, c, d \},
5,
\{ b, c, e \},
2,
\{ b, c, e \}
\end{align*}
\end{example}

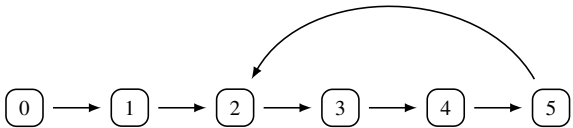
\begin{tikzpicture}[scale=0.70]
    \node (Box0) at (0,2) {\boxednode{0}};
    \node (Box1) at (2,2) {\boxednode{1}};
    \node (Box2) at (4,2) {\boxednode{2}};
    \node (Box3) at (6,2) {\boxednode{3}};
    \node (Box4) at (8,2) {\boxednode{4}};
    \node (Box5) at (10,2) {\boxednode{5}};
    
    \draw[semithick, ->] (Box0) to (Box1);
    \draw[semithick, ->] (Box1) to (Box2);
    \draw[semithick, ->] (Box2) to (Box3);
    \draw[semithick, ->] (Box3) to (Box4);
    \draw[semithick, ->] (Box4) to (Box5);
    \draw[semithick, ->, bend right=60] (Box5) to (Box2);
\end{tikzpicture}

\begin{definition}
  We say a CFG trace $S_0, k_0, \dots$ has $n$ loops
  iff
  there exists some index $k$ such that $k$ appears exactly $n+1$ times
  in the trace, and no other index that
  appears more than $n+1$ times in
  the trace.
\end{definition}

\begin{lemma} \label{main_lemma0}
  For any CFG $\langle V, E \rangle$,
  if all well-formed CFG traces up to 2 loops are valid,
  then all well-formed CFG traces are valid.
\end{lemma}
{\setlength{\parindent}{0pt}
  \textbf{Proof}
  
  We pick arbitrary $\langle V, E \rangle$,
  and we assume all well-formed traces of $\langle V, E \rangle$
  up to 2 loops are valid.
  
  We now want to show that
  for all number of loops $n$,
  for all trace $\tau$,
  if $\tau$ has up to $n$ loops,
  then $\tau$ is valid.
  
  We prove by induction over number
  of loops $n$.
  
  \textbf{Base case}, $n \leq 2$ loops: property holds by assumption.
  
  \textbf{Step case}:
  \begin{adjustwidth}{1em}{}
    We pick arbitrary $n \geq 2$.
    
    Induction hypothesis (IH1):
    for all trace $\tau$,
    if $\tau$ has up to $n$ loops,
    then $\tau$ is valid.

    We wish to prove that any trace
    with up to $n+1$ loops is also valid.
    Note that there are two cases where we yield a trace
    with $n+1$ loops:
    
    \begin{itemize}
    	\item
    	One is we add a suffix to a trace with $n$ loops
    	such that the suffix increases the number of
    	occurrence of some index from $n+1$ to $n+2$.
    	In this case, we can apply IH1.
    	
    	\item
    	The other case is we add a suffix to a trace with $n+1$ loops,
    	but the suffix does not increase the occurance
    	of any index to beyond $n+2$.
    	In this case, we cannot apply IH1.
    \end{itemize}
    
    As such we prove by induction over the number
    of indices with $n+2$ occurrence in $\tau'$.
    
    \textbf{Base case}, there are $m = 0$ indices which occurs $n+2$ times:
    \begin{adjustwidth}{1em}{}
      
      We pick arbitrary trace $\tau'$ with 0 indices which occur $n+2$ times.
      This follows directly from IH1 as $\tau'$ has up to $n$ loops.
    \end{adjustwidth}
  
    \textbf{Step case}:
    \begin{adjustwidth}{1em}{}
      
      We pick arbitrary $m > 0$.
      
      Induction hypothesis (IH2): for all trace $\tau$ up to $n+1$ loops,
      if $\tau$ has up to $m$ indices which
      appear $n+2$ times,
      then $\tau$ is valid.
      
      We then prove for arbitrary trace $\tau'$ which has $n+1$ loops,
      and up to $m+1$ indices which occurs $n+2$ times, $\tau'$ is valid.
      
      Then $\tau'$ is of the form
      $S_0, P, k, W, k, X, k, Y, k, Z$,
      for some initial state $S_0$,
      some prefix $P$,
      some rule $k$ that appears $n+2$ times,
      and some sequence
      $W$, $X$, $Y$, $Z$.
      
      We also let
      $W_{ids}$, $X_{ids}$, $Y_{ids}$ and $Z_{ids}$
      such that:
      \begin{align*}
        k, W &= tr(S_{end}(P), W_{ids})\\
        k, X &= tr(S_{end}(W), X_{ids})\\
        k, Y &= tr(S_{end}(X), Y_{ids})\\
        k, Z &= tr(S_{end}(Y), Z_{ids})
      \end{align*}
      
      By IH2, the following traces are valid:
      \begin{itemize}
        \item $S_0, P, k, W, k, X, k, Y_1$
        \item $S_0, P, k, W, k, X, k, Z_1$
        \item $S_0, P, k, W, k, Y_2, k, Z_2$
      \end{itemize}
      
      where
      \begin{align*}
        Y_1 &= tr(S_{end}(X), Y_{ids})\\
        Z_1 &= tr(S_{end}(X), Z_{ids})\\
        Y_2 &= tr(S_{end}(W), Y_{ids})\\
        Z_2 &= tr(S_{end}(Y_2), Z_{ids})
      \end{align*}
      
      By Lemma~\ref{lemma5}, trace $\tau'$ is also valid.
      
    \end{adjustwidth}
  \end{adjustwidth}

  $\square$
}

\begin{definition}
  The \emph{context}, or the set of defined cells, required for CFG trace $k_0, S_1, \dots, S_n, k_n, S_{n+1}$ is defined as follows:
  For all cell $c$,
  \begin{align*}
    c \in ctx_r(k_0, S_1, \dots, S_n, k_n, S_{n+1})
  \end{align*}
  if and only if
  \begin{align*}
    c \in cu_r(k_0)
    \lor (c\in ctx_r(k_1, \dots, S_{n+1}) \land c\notin cu_d(k_0)).
  \end{align*}
\end{definition}

\begin{definition} \label{def:ctx_maxR}
  We define ``maximal context required'' of a rule $k$ as
  $ctx_{maxR}(k) = \bigcup_{i=0}^n ctx_r(\tau^0_i)$,
  where $\{ \tau^0_0, \dots, \tau^0_n \}$ is the set of all CFG traces
  that begin with $k$ and are loop free.
\end{definition}

\begin{lemma} \label{ctx_maxR_is_max}
  For all CFG trace $\tau$,
  for all rule $k$, 
  if $\tau$ starts with $k$,
  then $ctx_r(\tau) \subseteq ctx_{maxR}(k)$.
  
  This reads: the ``maximal context'' we defined
  above is indeed ``maximal'' even in presence of loops.
\end{lemma}
{\setlength{\parindent}{0pt}
  \textbf{Proof}
  
  We pick arbitrary trace $\tau$.
  
  We want to show that
  if $\tau$ starts with $k$ for some $k$,
  $ctx_r(\tau) \subseteq ctx_{maxR}(k) = \bigcup_{i=0}^m ctx_r(\tau^0_i)$,
  where $\{ \tau^0_0, \dots, \tau^0_m \}$ is the set of all traces
  that begin with $k$ and is loop free.
  
  We prove by induction over number of
  loops $\tau$ has.
  
  \textbf{Base case}, $\tau$ is loop free: property holds trivially.
  
  \textbf{Step case}:
  \begin{adjustwidth}{1em}{}
    We pick arbitrary $n$.
    
    Induction hypothesis (IH1): for all trace $\tau$,
    for all $k$, if $\tau$ starts with $k$ and has up to $n$ loops,
    then
    $ctx_r(\tau) \subseteq ctx_{maxR}(k)$.
    
    Now we want to show for all trace $\tau$,
    for all $k$, if $\tau$ starts with $k$ and has $n+1$ loops,
    $ctx_r(\tau) \subseteq ctx_{maxR}(k)$
    
    With similar reasoning to
    proof of Lemma~\ref{main_lemma0},
    we prove by induction over the number of indices
    with $n+2$ occurrence in $\tau'$.
    
    \textbf{Base case}, there are $m = 0$ indices which occur $n+2$ times:
    \begin{adjustwidth}{1em}{}
    	We pick arbitrary trace $\tau'$ with 0 indices which occur $n+2$ times.
    	This follows directly from IH1 as $\tau'$ has up to $n$ loops.
    \end{adjustwidth}
  
  	\textbf{Step case}:
  	\begin{adjustwidth}{1em}{}
  		We pick arbitrary $m > 0$.
  		
  		Induction hypothesis (IH2): for all trace $\tau$,
  		for all $k$,
  		if $\tau$ starts with $k$ and has up to $n+1$ loops
  		and $\tau$ has up to $m$ indices which appear $n+2$ times,
  		then $ctx_r(\tau) \subseteq ctx_{maxR}(k)$.
  		
  		Now we wish to show this is also the case for any trace up to $n+1$ loops
  		with up to $m+1$ indices which appear $n+2$ times.
  		
  		Suppose not, then we have trace $\tau'$ which starts with some $k$,
  	  has $n+1$ loops,
  	  and has $m+1$ indices which appear $n+2$ times,
  		such that
  		$ctx_r(\tau') \not \subseteq ctx_{maxR}(k)$,
  		i.e. there exists cell $a \in ctx_r(\tau')$
  		and $a \notin ctx_{maxR}(k)$.
  		
  		Since $\tau'$ starts with $k$ and has $n+1$ loops,
  		$\tau'$ has the form $k, X, k', Y, k', Z$
  		where $k'$ appears $n+2$ times for some rule $k'$,
  		for some sequences $X$, $Y$, and $Z$.
  		
  		Then we have that trace $\tau''_1 = k, X, k', Y$
  		and trace $\tau''_2 = k, X, k', Z$,
  		both have $m$ indices which appear $n+2$ times.
  		
  		By IH2, $ctx_r(\tau''_1) \subseteq ctx_{maxR}(k)$,
  		as such we have $a \notin ctx_r(\tau''_1)$.
  		Also by IH2, $ctx_r(\tau''_2) \subseteq ctx_{maxR}(k)$,
  		as such we have $a \notin ctx_r(\tau''_2)$.
  		
  		However, by Lemma~\ref{lemma_ctx0},
  		$ctx_r(\tau') \subseteq ctx_r(\tau''_1) \cup ctx_r(\tau''_2)$.
  		As such $a \notin ctx_r(\tau')$, contradiction.
  	\end{adjustwidth}
  \end{adjustwidth}
	$\square$
}

\begin{definition} \label{def:ctx_maxRA}
  Given a CFG $\langle V, E \rangle$
  the ``maximal context required afterwards''
  of a rule $k$ is defined as
  $ctx_{maxRA}(k) = \bigcup_{i=0}^n ctx_{maxR}(k_i)$
  where $\{ k_0, \dots, k_n \} = succ(E, k).$ 
\end{definition}

\begin{lemma} \label{lemma_ctx0}
	For all $P_{ids}$, $X_{ids}$, $Y_{ids}$,
	\begin{align*}
		ctx_r([P, X]) \cup ctx_r([P, Y_1]) \supseteq
		ctx_r([P, X, Y_2])
	\end{align*}
	where
	\begin{align*}
		P &= tr(S_0, P_{ids})\\
		X &= tr(S_{end}(P), X_{ids})\\
		Y_1 &= tr(S_{end}(P), [k, Y_{ids}])\\
		Y_2 &= tr(S_{end}(X), [k, Y_{ids}])
	\end{align*}
\end{lemma}

{\setlength{\parindent}{0pt}
	\textbf{Proof}
	
	We pick arbitrary $Y_{ids}$.
	
	We prove by induction over $X$ (in reverse).
	
	\textbf{Base case}, $X$ is empty: the property holds trivially.
	
	\textbf{Step case}, $X = [x_0, S_1, \dots, x_n, S_{n+1}]$ for some $n$:
	\begin{adjustwidth}{1em}{}
		Induction hypothesis (IH1):
		$\forall P. ctx_r([P, x_1, S_2, \dots, S_{n+1}]) \cup ctx_r([P, Y_1])
		\supseteq
		ctx_r([P, x_1, S_2, \dots, S_{n+1}, Y_2])$.
		
		We wish to show that
		$\forall P. ctx_r([P, x_0, S_1, \dots, S_{n+1}]) \cup ctx_r([P, Y_1])
		\supseteq
		ctx_r([P, x_0, S_1, \dots, S_{n+1}, Y_2])$.
		
		We prove by induction over $P$.
		
		\textbf{Base case}, $P$ is empty:
		\begin{adjustwidth}{1em}{}
			Suppose the property does not hold,
			i.e.
			$ctx_r([x_0, S_1, \dots, x_n, S_{n+1}]) \cup ctx_r(Y_1)
			\not\supseteq ctx_r([x_0, S_1, \dots, x_n, S_{n+1}, Y_2])$,
			then there exists cell $a$ such that
			$a \notin ctx_r([x_0, S_1, \dots, x_n, S_{n+1}])
			\cup
			ctx_r(Y_1)$
			and
			$a \in ctx_r([x_0, S_1, \dots, x_n, S_{n+1}, Y_2])$.
			
			\textbf{Case} $a \notin cu_r(x_0)$:
			\begin{adjustwidth}{1em}{}
				\textbf{Case} $a \notin cu_d(x_0)$:
				\begin{adjustwidth}{1em}{}
					Then we have $a \in ctx_r([x_1, S_2, \dots, S_{n+1}])$.
					By IH1, we have that $a \in
					ctx_r([x_1, S_2, \dots, x_n, S_{n+1}]) \cup ctx_r(Y_1)
					\subseteq
					ctx_r([x_0, S_1, \dots, x_n, S_{n+1}]) \cup ctx_r(Y_1)
					$, contradiction.
				\end{adjustwidth}
				
				\textbf{Case} $a \in cu_d(x_0)$:
				\begin{adjustwidth}{1em}{}
					Then $a \notin ctx_r([x_0, S_1, \dots, x_n, S_{n+1}, Y_2])$, contradiction.
				\end{adjustwidth}
			\end{adjustwidth}
			
			\textbf{Case} $a \in cu_r(x_0)$:
			\begin{adjustwidth}{1em}{}
				Then we have $a \in ctx_r([x_0, S_1, \dots, x_n, S_{n+1}])
				\subseteq
				ctx_r([x_0, S_1, \dots, x_n, S_{n+1}]) \cup ctx_r(Y_1)$,
				contradiction.
			\end{adjustwidth}
		\end{adjustwidth}
	
		\textbf{Step case}, $P = [p_0, \dots, p_m, S_{m+1}]$:
		\begin{adjustwidth}{1em}{}
			Induction hypothesis (IH2):
			
			$ctx_r([p_1, \dots, p_m, S_{m+1}, x_0, S_1, \dots, S_{n+1}])
			\cup ctx_r([p_1, \dots, p_m, S_{m+1}, Y_1])
			\supseteq
			ctx_r([p_1, \dots, p_m, S_{m+1}, x_0, S_1, \dots, S_{n+1}, Y_2])$.
			
			Suppose the property does not hold,
			i.e.
			
			$ctx_r([p_0, \dots, p_m, S_{m+1}, x_0, S_1, \dots, x_n, S_{n+1}]) \cup ctx_r(Y_1)
			\not\supseteq ctx_r([p_0, \dots, p_m, S_{m+1}, x_0, S_1, \dots, x_n, S_{n+1}, Y_2])$,
			then there exists cell $a$ such that
			$a \notin ctx_r([P_0, \dots, p_m, S_{m+1}, x_0, S_1, \dots, x_n, S_{n+1}])
			\cup
			ctx_r([p_0, \dots, p_m, S_{m+1}, Y_1])$
			and
			$a \in ctx_r([p_0, \dots, p_m, S_{m+1}, x_0, S_1, \dots, x_n, S_{n+1}, Y_2])$.
			
			\textbf{Case} $a\notin cu_r(p_0)$:
			\begin{adjustwidth}{1em}{}
				\textbf{Case} $a\notin cu_d(p_0)$:
				\begin{adjustwidth}{1em}{}
					Then we have $a\in ctx_r([p_1, \dots, p_m, S_{m+1}, x_0, S_1, \dots, x_n, S_{n+1}, Y_2])$.
					By IH1, we have that $a \in
					ctx_r([p_1, \dots, p_m, S_{m+1}, x_1, S_2, \dots, x_n, S_{n+1}]) \cup ctx_r(Y_1)
					\subseteq
					ctx_r([p_0, \dots, p_m, S_{m+1}, x_0, S_1, \dots, x_n, S_{n+1}]) \cup ctx_r(Y_1)
					$, contradiction.
				\end{adjustwidth}
			
				\textbf{Case} $a\in cu_d(p_0)$:
				\begin{adjustwidth}{1em}{}
					Then $a \notin ctx_r([p_0, \dots, p_m, S_{m+1}, x_0, S_1, \dots, x_n, S_{n+1}, Y_2])$, contradiction.
				\end{adjustwidth}
			\end{adjustwidth}
		
			\textbf{Case} $a\in cu_r(p_0)$:
			\begin{adjustwidth}{1em}{}
				Then we have
				$a \in ctx_r([p_0, S_1, \dots, x_n, S_{n+1}])
				\subseteq
				ctx_r([p_0, S_1, \dots, x_n, S_{n+1}]) \cup ctx_r([Y_1])$,
				contradiction.
			\end{adjustwidth}
		\end{adjustwidth}
	\end{adjustwidth}
	
	$\square$
}

\section{Soundness and completeness}

We first relate the inferred context to possible
state of process memory in the following lemmas.
This is to show that the
approximation of memory content used in
translations of both
forward and backward exit biased rules
is correct.

\begin{lemma} \label{ctx_and_memory0}
	If all CFG traces are valid,
	then for all execution trace
	$(S_{tg0}, K_{tg0}, M_{tg0}), l_{tg0}\inbrac a_{tg0} \outbrac r_{tg0}, \dots$,
	for all $i$, for all $id$, \textbf{if}
	\begin{align*}
		id \in \mathbb{D}(K_{tgi})
		\land
		id \in \mathbb{D}(M_{tgi})
	\end{align*}
	\textbf{then}
    \begin{align*}
        ctx_{maxR}(K_{tgi}(id)) &\subseteq \mathbb{D}(M_{tgi}(id))  & \text{(1)} \\
       ctx_{maxRA}(K_{tgi}(id)) &\subseteq \mathbb{D}(M_{tgi+1}(id))  & \text{(2)}
    \end{align*}
\end{lemma}

\phantomsection
\label{proof_of_ctx_and_memory0}
{\setlength{\parindent}{0pt}
	\textbf{Proof of (1)}
	
	We pick arbitrary trace $\tau$ induction over size of $\tau$.
	
	\textbf{Base case}, $\tau$ is empty: the property holds vacuously.
	
	\textbf{Step case}, $\tau = (S_{tg0}, K_{tg0}, M_{tg0}), l_{tg0}\inbrac a_{tg0} \outbrac r_{tg0}, \dots, (S_{tgn+1}, K_{tgn+1}, M_{tgn+1})$:
	\begin{adjustwidth}{1em}{}
		IH: for all $i \leq n$, $id$, if $id \in \mathbb{D}(K_{tgi}) \land id \in \mathbb{D}(M_{tgi})$, then $\mathbb{D}(M_{tgi}(id)) \supseteq ctx_{maxR}(K_{tgi}(id))$
		
		We show the property also holds for $i = n+1$.
		
		We case analyze $(S_{tgn}, K_{tgn}, M_{tgn}), l_{tgn}\inbrac a_{tgn} \outbrac r_{tgn}, (S_{tgn+1}, K_{tgn+1}, M_{tgn+1})$ by semantic rule used:
		\begin{adjustwidth}{1em}{}
			\textbf{Case} \textsc{Fresh} or \textsc{Start}: the property holds trivially.
			
			\textbf{Case} \textsc{Rule}:
			\begin{adjustwidth}{1em}{}
				Then we have some $id$ where
				$id \in \mathbb{D}(K_{tgn})$
				and
				$id \in \mathbb{D}(M_{tgn})$.
				
				Suppose the property does not hold, then we have cell $a$
				such that
				(1) $a \notin \mathbb{D}(M_{tgn+1}(id))$
				and
				(2) $a\in ctx_{maxR}(K_{tgn+1}(id))$.
				
				Since (1) $a \notin \mathbb{D}(M_{tgn+1}(id))$,
				we know (3) $a \notin cu_d(K_{tgn}(id))$ by IH.
				
				Since (2) $a\in ctx_{maxR}(K_{tgn+1}(id))$ and (3) $a \notin cu_d(K_{tgn}(id))$,
				by definition, (4) $a \in ctx_{maxR}(K_{tgn}(id))$.
				Then by IH, we also have (5) $a \in \mathbb{D}(M_{tgn}(id))$.
				
				We then case analyze cell usage with respect to $a$:
				
				\textbf{Case} $a\notin cu_u(K_{tgn}(id))$:
				\begin{adjustwidth}{1em}{}
					By (5), we have $a \in \mathbb{D}(M_{tgn+1}(id))$, contradiction to (1).
				\end{adjustwidth}

				\textbf{Case} $a\in cu_u(K_{tgn}(id))$:
				\begin{adjustwidth}{1em}{}
					By (2), there exists CFG trace $\tau' = K_{tgn+1}(id), \dots$
					such that
					$a \in ctx_r(\tau')$.
					But since $a \in cu_u(K_{tgn}(id))$,
					we have invalid CFG trace $S, K_{tgn}(id), S', \tau'$,
					for some $S$, $S'$,
					contradicting our assumption that all CFG traces are valid.
				\end{adjustwidth}
			\end{adjustwidth}
			
		\end{adjustwidth}
	\end{adjustwidth}

	$\square$
}

\phantomsection
{\setlength{\parindent}{0pt}
  \textbf{Proof of (2)}
  
  This follows directly from Definition~\ref{def:ctx_maxRA} (definition of $ctx_{maxRA}$) and Lemma~\ref{ctx_and_memory0}.
  
  $\square$
}

We now state what
Tamarin trace corresponding to a Tamgram trace
means in two parts:

\begin{definition}
  Tamarin trace $S_{t0}, l_{t0} \inbrac a_{t0} \outbrac r_{t0}, \dots, S_{tn}$
  is \textbf{state consistent} w.r.t. Tamgram trace
  $(S_{tg0}, K_{tg0}, M_{tg0}), l_{tg0} \inbrac a_{tg0} \outbrac r_{tg0}, \dots, (S_{tgn}, K_{tgn}, M_{tgn})$
  if and only if $\forall i. $
  \begin{align*}
    S_{ti} = \bigcup_{id=id_0}^{id_j}\{ St^*_i (id, k_i, deref(m_i, ctx_i)) \} \cup^\sharp S_{tgi}
  \end{align*}
  where
  \begin{align*}
    \langle St^*_i, k_i, ctx_i \rangle &= \langle St^F, k, ctx_{maxR}(k_i) \rangle\\
                   & \phantom{= } \text{or } \langle St^B, k', ctx_{maxRA}(k') \rangle\\
    k &= K_{tgi}(id)\\
    k' &\in pred(E, k)\\
    m_i &= M_{tgi}(id)\\
    \{ id_0, \dots, id_j \} &= \mathbb{D}(K_{tgi}) = \mathbb{D}(M_{tgi})
  \end{align*}
\end{definition}

\begin{definition}
  Tamarin trace $S_{t0}, l_{t0} \inbrac a_{t0} \outbrac r_{t0}, \dots, S_{tn}$
  is \textbf{step consistent} w.r.t. to Tamgram trace
  $(S_{tg0}, K_{tg0}, M_{tg0}), l_{tg0} \inbrac a_{tg0} \outbrac r_{tg0}, \dots, (S_{tgn}, K_{tgn}, M_{tgn})$
  if and only if $\forall i. \exists id.$ conjunction of the following holds:
  \begin{align}
  	lfacts(S_{ti}) &\ni St^*_i(id, k_i, deref(m_i, ctx_i)) \\
  	l_{ti} &= \{ St^*_i(id, k_i, deref(m_i, ctx_i)) \} \\
    &\phantom{= } \cup^\sharp deref(m_i, l_{tgi}) \nonumber \\
  	a_{ti} &= deref(m_i, a_{tgi})\\
  	r_{ti} &= \{ St^*_{i+1}(id, k_{i+1}, deref(m_{i+1}, ctx_{i+1})) \}\\
    &\phantom{= } \cup^\sharp deref(m_i, nostmt(r_{tgi})) \nonumber \\
  	lfacts(S_{ti+1}) &\ni St^*_{i+1}(id, k_{i+1}, deref(m_{i+1}, ctx_{i+1}))
  \end{align}
  where
  \begin{align*}
    \langle St^*_i, k_i, ctx_i \rangle &= \langle St^F, k \rangle \text{ or } \langle St^B, k' \rangle, \\
    \langle St^*_{i+1}, k_{i+1}, ctx_{i+1} \rangle &= \langle St^F, k'' \rangle \text{ or }\langle St^B, k \rangle, \\
    k = K_{tgi}(id),  & \quad k' \in pred(k), \quad   
    k'' \in succ(k), \\
    m_i = M_{tgi}(id) & \quad  m_{i+1} = M_{tgi+1}(id), \\  
    set(ctx_i) \subseteq \mathbb{D}(m_i), & \quad 
    set(ctx_{i+1}) \subseteq \mathbb{D}({m_{i+1}}).
  \end{align*}
\end{definition}

\begin{definition}
  A Tamarin trace $\tau_{t}$ corresponds to Tamgram trace $\tau_{tg}$
  if and only if
  $\tau_{t}$ is both state consistent and step consistent
  with respect to $\tau_{tg}$.
\end{definition}

\begin{theorem}[Soundness]
  For all Tamgram system $Sys$,
  for all Tamgram trace $(S_{tg0}, K_{tg0}, M_{tg0}), ru_{tg0}, \dots, (S_{tgn}, K_{tgn}, M_{tgn})$
  of $Sys$,
  there exists Tamarin trace $S_{t0}, ru_{t0}, \dots, S_{tn}$ of $T(Sys)$
  which corresponds to the Tamgram trace.
\end{theorem}

{\setlength{\parindent}{0pt}
  \textbf{Proof}
  
We pick arbitrary Tamgram system $Sys$.

We prove by induction over the length of Tamgram trace.

\textbf{Base case}, Tamgram trace is empty, the property holds vacuously.

\textbf{Step case},
Tamgram trace is
$(S_{tg0}, K_{tg0}, M_{tg0}), ru_{tg0}, \dots,
(S_{tgn+1}, K_{tgn+1}, M_{tgn+1})$.
\begin{adjustwidth}{1em}{}
        IH: for subtrace
    $(S_{tg0}, K_{tg0}, M_{tg0}), ru_{tg0}, \dots,
    (S_{tgn}, K_{tgn}, M_{tgn})$,
    there exists a Tamarin trace
    $S_{t0}, ru_{t0}, \dots, S_{tn}$
    that corresponds to the subtrace.
    
    We now case analyse
    $(S_{tgn}, K_{tgn}, M_{tgn}),$
    $l_{tgn}\inbrac a_{tgn} \outbrac r_{tgn},$
    $(S_{tgn+1}, K_{tgn+1}, M_{tgn+1})$
    by semantic rule used,
    and show that we can yield the final
    Tamarin trace $S_{t0}, ru_{t0}, \dots, S_{tn}, ru_{tn}, S_{tn+1}$
    by picking the right translated rule to use for $ru_{tn}, S_{tn+1}$.
    
    If the step is picked by semantic rule \textsc{Fresh},
    then we can use the rule \textsc{Fresh}.
    
    If the step is picked by semantic rule \textsc{Start},
    then we can use the starting
    rule added by $cfg$ (Definition~\ref{def-cfg-fun}).
    We can check that in both cases
    the property holds trivially.
    
    For case of step being picked by semantic rule \textsc{Rule}:
    \begin{adjustwidth}{1em}{}
        \textit{\textbf{To show the corresponding translated rule exists in
        the translated system}}:
        \begin{adjustwidth}{1em}{}
            Since $(S_{tgn}, K_{tgn}, M_{tgn}),$
            $l_{tgn}\inbrac a_{tgn} \outbrac r_{tgn},$
            $(S_{tgn+1}, K_{tgn+1}, M_{tgn+1})$
            is a step in the Tamgram trace,
            we know
            there exists vertex
            $(k_{n}, l \inbrac~a~\outbrac r) \in V$
            and edge
            $(k_{n}, k_{n+1}) \in E$
            such that
            $l_{tgn} = \phi(l)$,
            $a_{tgn} = \phi(a)$,
            $r_{tgn} = \phi(r)$,
            where $\phi$ is a ground substitution replacing
            all schematic variables with ground terms.
            This means the corresponding translated rule
            $[St^*_n(\sim pid, k_{tn}, cvar(ctx_n)), cvar(l)]$
            $\inbrac cvar(a) \outbrac$
            $[St^*_{n+1}(\sim pid, k_{tn+1}, cvar(ctx_{n+1})), cvar(nostmt(r))]$
            exists
            in the translated system,
            for some $ctx_n$, $ctx_{n+1}$,
            $St^*_n = St^F$ or $St^B$, $St^*_{n+1} = St^F$ or $St^B$,
            where $k_{tn} = k_{n}$ or $\in pred(E, k_{n})$,
            $k_{tn+1} = k_{n}$ or $\in succ(E, k_{n})$.
        \end{adjustwidth}
        
        \textit{\textbf{To show we can construct a ground instance
        of said translated rule}}:
        \begin{adjustwidth}{1em}{}
            Since cell dereference is exhaustive by \textsc{Rule},
            and by Lemma~\ref{ctx_and_memory0},
            process memory contains enough content whether $ctx_{maxR}$ or $ctx_{maxRA}$ is used
            during translation,
            we can define $\phi' = \phi \cup \{ cvar(c_0) \mapsto x_0, \dots \}$
            where
            $\{ c_0 \mapsto x, \dots \} = M_{tgn}(id)$
            such that
            we have ground instance of the translated rule
            $\phi'(cvar([St^*_n(\dots), l] \inbrac a \outbrac [St^*_{n+1}(\dots), nostmt(r)]))$
            which is equal to
            $\phi(deref(M_{tgn}(id), [St^*_n(\dots), l] \inbrac a \outbrac [St^*_{n+1}(\dots), nostmt(r)]))$.
        \end{adjustwidth}
        
        \textit{\textbf{To show left side of the translated rule
        $\phi'(cvar([St^*_n(\dots), l]))$ can be fulfilled}}:
        \begin{adjustwidth}{1em}{}
            By IH and state consistency, $\phi(cvar(St^*_n(\dots))) \in lfacts(S_{tn})$.
            By \textsc{Rule}, we know $deref(M_{tgn}(id), \phi(l))
            = \phi'(cvar(l)) \subseteq^+ S_{tgn}$.
            By IH and state consistency again, $S_{tgn} \subseteq^+ S_{tn}$.
            Overall $\phi'(cvar([St^*_n(\dots), l])) \subseteq^+ S_{tn}$.
            
            Above shows we can make a move
            using the translated rule.
            We use similar reasoning and refer to the translation procedure
            to show the resulting move retains state and step consistency.
        \end{adjustwidth}
    \end{adjustwidth}
\end{adjustwidth}
}

\begin{theorem}[Completeness]
  For all Tamgram system $Sys$,
  for all Tamarin trace $S_{t0}, ru_{t0}, \dots, S_{tn}$ of $T(Sys)$,
  there exists Tamgram trace $(S_{tg0}, K_{tg0}, M_{tg0}), ru_{tg0}, \dots, (S_{tgn}, K_{tgn}, M_{tgn})$
  of $Sys$,
  which the Tamarin trace corresponds to.
\end{theorem}

{\setlength{\parindent}{0pt}
  \textbf{Proof}
  
We pick arbitrary Tamgram system $Sys$.

We prove by induction over the length of Tamarin trace of
the translated system $T(Sys)$.

\textbf{Base case}, Tamarin trace is empty, the property holds vacuously.

\textbf{Step case}, Tamarin trace is $S_{t0}, ru_{t0}, \dots, S_{tn}, ru_{tn}, S_{n+1}$.
\begin{adjustwidth}{1em}{}
    IH: is then subtrace
    $S_{t0}, ru_{t0}, \dots, S_{tn}$
    corresponds to some Tamgram trace
    $(S_{tg0}, K_{tg0}, M_{tg0}), ru_{tg0}, \dots, (S_{tgn}, K_{tgn}, M_{tgn})$.
    
    We now case analyse $S_{tn}, ru_{tn}, S_{tn+1}$ by translated rule used.
    
    If the step is picked by FRESH,
    then we can use the counterpart on Tamgram's side.
    
    If the step is picked by a rule
    added by procedure $cfg$ (Definition~\ref{def-cfg-fun}),
    we can copy the move using semantic rule \textsc{Start},
    and we can check that the property holds trivially.
    
    For case of step being picked by any other rule:
    
    \textit{\textbf{To show that we can retrieve the original rule}}:
    \begin{adjustwidth}{1em}{}
        Since $S_{tn}, ru_{tn}, S_{tn+1}$
        is a step in the Tamarin trace,
        we know there exists vertex $(k_n, l\inbrac~a~\outbrac r) \in V$
        and edge $(k_n, k_{n+1}) \in E$
        such that $l_{tn} = \phi(St^*_{n}(\sim~pid, k_{tn}, cvar(ctx_n))), \phi(cvar(l))$,
        $a_{tn} = \phi(cvar(a))$,
        and
        $r_{tn} = \phi(St^*_{n+1}(\sim pid, k_{tn+1}, cvar(ctx_{n+1}))), \phi(cvar(nostmt(r)))$,
        for some $St^*_n = St^F$ or $St^B$,
        $St^*_{n+1} = St^F$ or $St^B$,
        $k_{tn} = k_{n}$ or $\in pred(E, k_n)$,
        $k_{tn+1} = k_{n}$ or $\in succ(E, k_n)$
        where $\phi$ is a ground substitution replacing all
        schematic variables with ground terms.
    \end{adjustwidth}
    
    \textit{\textbf{To show we can construct a ground instance
    of the original rule}}:
    \begin{adjustwidth}{1em}{}
        Since the translated rule carries
        more schematic variables than the original rule by
        the translation procedure, $\phi$ suffices
        for construction of the required ground instance.
    \end{adjustwidth}
    
    \textit{\textbf{To show the left side of the original rule
     can be fulfilled}}:
    \begin{adjustwidth}{1em}{}
        By IH and state consistency,
        $id = \phi(pid) \in \mathbb{D}(M_{tgn})$
        and $\in \mathbb{D}(K_{tgn})$.
        By Lemma~\ref{ctx_and_memory0},
        we know process memory $m_{tgn} = M_{tgn}(id)$ contains
        enough content to dereference cells in $l$.
        Since $deref(m_{tgn}, \phi(l)) \subseteq^\sharp l_{tn}$,
        $deref(m_{tgn}, \phi(l)) \subseteq^+ S_{tgn}$.
        
        Additionally, $m_{tgn}$ also carries
        enough content to dereference cells in $a$, $r$.
        We use similar reasoning to show the resulting
        move retains state and step consistency.
    \end{adjustwidth}
\end{adjustwidth}

}

\end{appendices}

\end{document}